# Suppression of spin bath and low-frequency noise for sub-MHz AC magnetometry based on double-dressed spin qubit in diamond

Kihwan Kim,[1] Yisoo Na,[1] Jungbae Yoon,[1] Dongkwon Lee,[1] Hee Seong Kang,[2] Chul-Ho Lee,[3] Chulki Kim,[4] and Donghun Lee[1,*]

[1]Department of Physics, Korea University, Seoul, 02841, Republic of Korea
[2]KU-KIST Graduate School of Converging Science and Technology, Korea University, Seoul, 02841, Republic of Korea
[3]Department of Electrical and Computer Engineering, Seoul National University, Seoul, 08826, Republic of Korea
[4]Quantum Information Center, Korea Institute of Science and Technology (KIST), Seoul, 02792, Republic of Korea

We experimentally demonstrate a protocol that effectively suppresses low-frequency noise, including the qubit–bath interaction in diamond. This enables the detection of AC signals below 1 MHz with an enhanced signal-to-noise ratio (SNR) of up to SNR $\approx$ 17. The method is based on AC magnetometry with single- and double-dressed states, which are adiabatically transferred from the initial bare qubit states via concatenated continuous dynamical decoupling (cCDD). We compare the dressed state protocols with a conventional pulsed dynamical decoupling (PDD) method, such as XY16. This work paves an alternative way toward the sensitive detection of weakly coupled nuclear spins in low-field NMR experiments and sub-MHz AC magnetometry.

## I. INTRODUCTION.

Some defects in solids have been proved to be useful resources in the field of quantum information science and technology [1, 2]. Among these defects, the negatively charged nitrogen-vacancy (NV) center in diamond has emerged as a leading system, particularly for quantum sensing applications, owing to its compatibility with high spatial resolution and high field sensitivity. One of the most promising applications is nuclear magnetic resonance (NMR) spectroscopy. Intensive studies on the NV-based NMR have been made employing diverse sets of quantum sensing methodologies, including pulsed dynamical decoupling (PDD) [3–9], correlation protocols [10–12], quantum memory [13–15], and synchronized readout [16–19].

Pioneering works have successfully demonstrated NMR sensitivity at the single-molecule level [20, 21] and sub-Hz spectral resolution [17–19]. However, the majority of the experiments rely on PDD techniques, which provide prolonged coherence time and enhanced sensitivity but are limited by finite pulse widths, pulse errors, and maximum power of the driving field [22, 23]. An alternative approach has been developed based on continuous driving of the qubits, known as continuous dynamical decoupling (CDD), with spin locking serving as a prominent example [24-37].

Spin locking is a technique for maintaining or locking the spin magnetization at a specific location on the Bloch sphere, often in the transverse plane, by continuously driving (dressing) the qubit. Spin locking was originally introduced in the long history of NMR [24, 25]. For instance, it was implemented to realize the Hartmann–Hahn condition between the spins of different energy scales – e.g., electron spins (GHz) and nuclear spins (MHz) – without usage of a large magnetic field [26]. Even in the absence of an external magnetic field, tunable control over the spin's energy levels is possible by adjusting the amplitude of the driving field. This provides not only selective tuning of the qubit energy over a wide frequency range, but also effective decoupling of the qubit from its noisy environment. Spin locking has been used for dynamical nuclear polarization (DNP) [26–32], zero-field electron paramagnetic resonance spectroscopy [33-35] and AC field sensing [36–38] at frequencies up to hundreds of MHz, which exceeds the detection bandwidth of traditional PDD approaches.

Recently, a more advanced form of CDD has been introduced: concatenated continuous driving, which applies a secondary driving field to a qubit [38–42]. This method, referred to as double-dressing, involves dressing the qubit twice with two continuous driving fields. It is expected to provide stronger noise decoupling and, therefore, enhanced sensitivity compared to the single dressing method [39–41]. Despite these theory proposals and early experimental demonstrations involving diamond NV centers, their application to quantum magnetometry, particularly in the low-frequency sub-MHz regime, has not been extensively studied.

In this paper, we show that the double dressing method can effectively suppress the interaction between the NV center and its nuclear spin bath, as well as low-frequency noise. This results in improved detection of sub-MHz magnetic signals. We use the concatenated CDD (cCDD) technique to adiabatically dress the spin qubit into single- and double-dressed states in succession.

First, we demonstrate that dressed qubits exhibits increased coherence time due to the combined effects of quenching magnetic noise, including spin bath and other noise sources. The Ramsey experiment shows that the effective dephasing time is extended by ~ 450 % for the single-dressed qubit and ~ 960 % for the double-dressed qubit compared to the bare qubit.

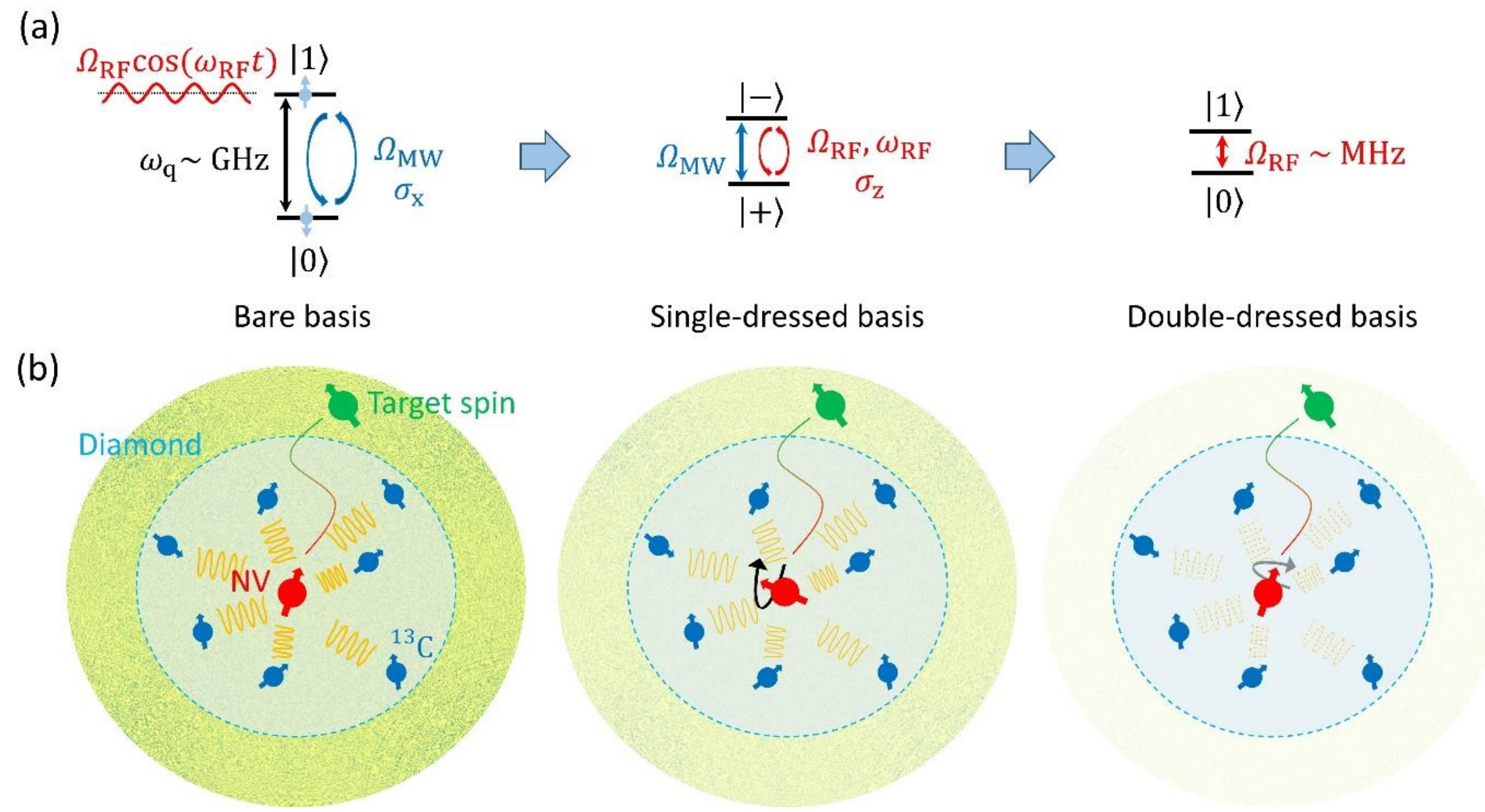

FIG 1. Illustration of double dressing method (a) Energy level diagrams for bare, single-, and double-dressed bases. Driven by continuous microwave (MW), the bare qubit transfers to a single-dressed qubit whose frequency is determined by the amplitude of the microwave field, $\Omega_{\mathrm{MW}}$. When the frequency of the radio-frequency (RF), $\omega_{\mathrm{RF}}$, matches with $\Omega_{\mathrm{MW}}$, the RF drive transfers the qubit to a double-dressed qubit with frequency, $\Omega_{\mathrm{RF}}$, which equals to the RF amplitude. (b) Along the transitions to the single- and double-dressed qubits, the hyperfine interaction between the NV center and $^{13}$C nuclear spin bath, as well as interactions with environmental noise, can be suppressed significantly more than the signal from target spins. This enables the detection of target signals, such as weakly coupled nuclear spins with enhanced signal-to-noise ratio (SNR).

Second, we show that the dressing methods substantially mitigate the spin bath coupling by reducing the spectral broadening and sub-MHz noise associated with $^{13}$C nuclear spin bath. As a result, $^{13}$C Larmor precession frequency is measured with enhanced SNR ($\approx 4$). We also conduct a test measurement using both stochastic and coherent signals at 0.8 MHz. With the double dressing method, we are able to achieve SNR $\approx 8$ for the stochastic source and SNR $\approx 17$ for the coherent source.

Finally, we test whether the dressing methods can suppress low-frequency noise beyond the spin bath, which may originate from the measurement system and the external environment. We evaluate the detection of a test tone at 0.6 MHz against added radio-frequency (RF) noise using the PDD, CDD, and cCDD techniques. From this comparison, we observe the advantage of using the cDD technique.

The double dressing method introduced in this paper effectively suppress both spin bath and low-frequency noise, offering potential applications in resilient qubit engineering, low-field NMR, and low-frequency AC magnetometry.

## II. THEORETICAL ANALYSIS OF SINGLE- AND DOUBLE-DRESSED QUBITS

The main idea of this paper is depicted in Fig. 1. We sequentially transfer the spin qubit from the bare basis to the single and double-dressed bases using two continuous driving fields. To simplify, we consider a two-level system with energy splitting of $\omega_{\mathrm{q}}$. The bare qubit Hamiltonian with spin 1/2 nuclear spin bath and RF noise can be written as:

$$H \approx \frac{\omega_{\mathrm{q}}}{2}\sigma_{\mathrm{z}} + \sigma_{\mathrm{z}}\sum_{\vec{\mathrm{r}}}\left[\frac{A_{||,\vec{\mathrm{r}}}}{2}I_{\mathrm{z},\vec{\mathrm{r}}} + \frac{A_{\perp,\vec{\mathrm{r}}}}{2}I_{\mathrm{x},\vec{\mathrm{r}}}\right] + \sum_{\vec{\mathrm{r}}}\gamma_{\mathrm{n},\vec{\mathrm{r}}}BI_{\mathrm{z},\vec{\mathrm{r}}} + \\ +\epsilon_{\mathrm{z}}\sigma_{\mathrm{z}} + \epsilon_{\mathrm{x}}\sigma_{\mathrm{x}} + \Omega_{\mathrm{MW}}\cos(\omega_{\mathrm{MW}}\mathrm{t})\,\sigma_{\mathrm{x}} + \Omega_{\mathrm{RF}}\cos(\omega_{\mathrm{RF}}\mathrm{t})\,\sigma_{\mathrm{z}} \quad (1)$$

where $\sigma_{\mathrm{x}}$, $\sigma_{\mathrm{y}}$, $\sigma_{\mathrm{z}}$ are the Pauli operators, $I_{\mathrm{z}}$, $I_{\mathrm{x}}$ are the nuclear spin operators with spin 1/2, $\vec{r}$ is the position vector of each nuclear spin, $\gamma_{\mathrm{n},\vec{\mathrm{r}}}$ is the nuclear spin gyromagnetic ratio, $B$ is the applied magnetic field, $A_{||,\vec{\mathrm{r}}}$ and $A_{\perp,\vec{\mathrm{r}}}$ are the parallel and perpendicular hyperfine coupling constants, $\epsilon_{\mathrm{z}}$ and $\epsilon_{\mathrm{x}}$ are RF noise along the z and x axes, and $\Omega_{\mathrm{MW}}$ and $\Omega_{\mathrm{RF}}$ are the amplitudes of the microwave (MW) and RF fields with frequencies $\omega_{\mathrm{MW}}$ and $\omega_{\mathrm{RF}}$, respectively. Note that we treat $\hbar$ as 1.

The first three terms in Eq. (1) denote the qubit energy, the hyperfine interaction between the electron spin and neighboring nuclear spins, and the nuclear spins' Zeeman energy. The following two terms describe the RF noise, while the final two terms correspond to the continuous driving fields (MW and RF) used for the transition to the single- and double-dressed states.

In the large detuning regime of $\delta_{\mathrm{MW}} = \omega_{\mathrm{q}} - \omega_{\mathrm{MW}} \gg A_{\perp,\vec{\mathrm{r}}},\ A_{||,\vec{\mathrm{r}}},\ \gamma_{\mathrm{n},\vec{\mathrm{r}}}B$, the hyperfine coupling and the nuclear Zeeman terms can be treated as small perturbations and the single-dressed qubit Hamiltonian after the first-order perturbation becomes

$$H_{\mathrm{SD}} \approx \frac{\omega_{\mathrm{SD}}}{2}\sigma_{\mathrm{z,SD}} + \cos(\theta_{\mathrm{SD}})\,\sigma_{\mathrm{z,SD}}\sum_{\vec{\mathrm{r}}}\left[\frac{A_{||,\vec{\mathrm{r}}}}{2}I_{\mathrm{z},\vec{\mathrm{r}}} + \frac{A_{\perp,\vec{\mathrm{r}}}}{2}I_{\mathrm{x},\vec{\mathrm{r}}}\right] + \sum_{\vec{\mathrm{r}}}\gamma_{\mathrm{n},\vec{\mathrm{r}}}BI_{\mathrm{z},\vec{\mathrm{r}}} + \sin(\theta_{\mathrm{SD}})\,\Omega_{\mathrm{RF}}\cos(\omega_{\mathrm{RF}}\mathrm{t})\,\sigma_{\mathrm{x,SD}} + \tilde{\epsilon}_{\mathrm{z}}\sigma_{\mathrm{z,SD}} + \tilde{\epsilon}_{\mathrm{x}}\sigma_{\mathrm{x,SD}} \quad (2)$$

where $\omega_{\mathrm{SD}} = \sqrt{\delta_{\mathrm{MW}}^2 + \Omega_{\mathrm{MW}}^2}$, $\tan(\theta_{\mathrm{SD}}) = \frac{\Omega_{\mathrm{MW}}}{\delta_{\mathrm{MW}}}$, $\sigma_{\mathrm{z,SD}} = \cos(\theta_{\mathrm{SD}})\,\sigma_{\mathrm{z}} + \sin(\theta_{\mathrm{SD}})\,\sigma_{\mathrm{x}}$, and $\sigma_{\mathrm{x,SD}} = \sin(\theta_{\mathrm{SD}})\,\sigma_{\mathrm{z}} - \cos(\theta_{\mathrm{SD}})\,\sigma_{\mathrm{x}}$, $\tilde{\epsilon}_{\mathrm{z}} = \cos(\theta_{\mathrm{SD}})\epsilon_{\mathrm{z}}$ and $\tilde{\epsilon}_{\mathrm{x}} = \sin(\theta_{\mathrm{SD}})\,\epsilon_{\mathrm{z}}$. Note that the transvers RF noise $\epsilon_{\mathrm{x}}$ is neglected due to the rotating wave approximation. Compared to the bare qubit, both the hyperfine couplings and RF noise are suppressed by a factor of $\cos(\theta_{\mathrm{SD}}) = \frac{\delta_{\mathrm{MW}}}{\omega_{\mathrm{SD}}}$.

We can follow the similar procedure for the double-dressing case. In the large detuning regime of $\delta_{\mathrm{RF}} = \omega_{\mathrm{SD}} - \omega_{\mathrm{RF}} \gg \frac{\delta_{\mathrm{MW}}}{\omega_{\mathrm{SD}}}A_{\perp,\vec{\mathrm{r}}},\ \frac{\delta_{\mathrm{MW}}}{\omega_{\mathrm{SD}}}A_{||,\vec{\mathrm{r}}}, \gamma_{\mathrm{n},\vec{\mathrm{r}}}B$, the double-dressed qubit Hamiltonian after the first-order perturbation becomes

$$H_{\mathrm{DD}} \approx \frac{\omega_{\mathrm{DD}}}{2}\sigma_{\mathrm{z,DD}} + \cos(\theta_{\mathrm{SD}})\cos(\theta_{\mathrm{DD}})\,\sigma_{\mathrm{z,DD}}\sum_{\vec{\mathrm{r}}}\left[\frac{A_{||,\vec{\mathrm{r}}}}{2}I_{\mathrm{z},\vec{\mathrm{r}}} + \frac{A_{\perp,\vec{\mathrm{r}}}}{2}I_{\mathrm{x},\vec{\mathrm{r}}}\right] + \sum_{\vec{\mathrm{r}}}\gamma_{\mathrm{n},\vec{\mathrm{r}}}BI_{\mathrm{z},\vec{\mathrm{r}}} + \tilde{\tilde{\epsilon}}_{\mathrm{z}}\sigma_{\mathrm{z,DD}} + \tilde{\tilde{\epsilon}}_{\mathrm{x}}\sigma_{\mathrm{x,DD}} \quad (3)$$

where $\omega_{\mathrm{DD}} = \sqrt{\delta_{\mathrm{RF}}^2 + \sin^2(\theta_{\mathrm{SD}})\,\Omega_{\mathrm{RF}}^2}$, $\tan(\theta_{\mathrm{DD}}) = \frac{\sin(\theta_{\mathrm{SD}})\Omega_{\mathrm{RF}}}{\delta_{\mathrm{RF}}}$, $\sigma_{\mathrm{z,DD}} = \cos(\theta_{\mathrm{DD}})\,\sigma_{\mathrm{z,SD}} + \sin(\theta_{\mathrm{DD}})\,\sigma_{\mathrm{x,SD}}$, $\sigma_{\mathrm{x,DD}} = \sin(\theta_{\mathrm{DD}})\,\sigma_{\mathrm{z,SD}} - \cos(\theta_{\mathrm{DD}})\,\sigma_{\mathrm{x,SD}}$, $\tilde{\tilde{\epsilon}}_{\mathrm{z}} = \cos(\theta_{\mathrm{DD}})\tilde{\epsilon}_{\mathrm{z}}$ and $\tilde{\tilde{\epsilon}}_{\mathrm{x}} = \sin(\theta_{\mathrm{DD}})\,\tilde{\epsilon}_{\mathrm{z}}$. In comparison to the bare qubit case, the hyperfine couplings and RF noise are further reduced by a total factor of $\cos(\theta_{\mathrm{SD}})\cos(\theta_{\mathrm{DD}}) = \frac{\delta_{\mathrm{MW}}}{\omega_{\mathrm{SD}}}\frac{\delta_{\mathrm{RF}}}{\omega_{\mathrm{DD}}}$. The effective suppression of the qubit-bath interaction and RF noise is the primary mechanism for probing the target signals, such as nuclear spins, with enhanced SNR, as illustrated in Fig. 1(b).

Note that we assume the large detuning regime to simplify the calculation. See Supplemental Material [43] for further discussions beyond the large detuning regime. In many cases, a small detuning is sufficient to improve the robustness of the dressed qubits, although a large detuning can provide advantages in certain quantum sensing experiments. This will be discussed further in section V.

## III. DRESSED STATE ELECTRON SPIN RESONANCE AND COHERENCE TIME

To demonstrate the suppression effect, we first performed electron spin resonance (ESR) and Ramsey interferometry experiments on each qubit of the bare, single-, and double-dressed bases, as shown in Fig. 2. Adiabatic passage is used to coherently transform each dressed eigenstate to the corresponding qubit eigenstate [44–47]. As illustrated in Fig. 2(a), we gradually changed the amplitude of the continuous driving fields over a slow ramping time. In order to do this within the time constraints of the qubit decoherence, we intentionally set a non-zero detuning to accelerate the passage. For instance, using finite detuning, we are able to shorten the ramping time by an order of magnitude while maintaining the same adiabatic state transfer as on-resonant ramping. A more detailed discussion on the adiabaticity is available in the Supplemental Material [43].

The upper graphs in Fig. 2(b)-(d) show the ESR results for qubits with three different bases. Here, we applied a ~ 400 G magnetic field to polarize the $^{15}$N nuclear spin into $m_{\mathrm{I}} = +1/2$ state. Note that nitrogen spin polarization is not an essential requirement of our sensing protocol. When it is not polarized, such as when the applied magnetic field is significantly smaller than 512 G, the hyperfine sublevels produce replica signal peaks as shown in [36], which can be eliminated by the methods introduced in [36].

The bare qubit frequency of 4,007 MHz corresponds to the NV transition between $m_{\mathrm{s}} = 0$ and $m_{\mathrm{s}} = +1$ spin states. The single-dressed qubit operates at 16.5 MHz, with the continuous drive at $\Omega_{\mathrm{MW}} = 16$ MHz and $\omega_{\mathrm{MW}} = 4{,}011$ MHz ($\delta_{\mathrm{MW}} = 4$ MHz). Similarly, the second field at $\Omega_{\mathrm{RF}} = 2$ MHz and $\omega_{\mathrm{RF}} = 15$ MHz ($\delta_{\mathrm{RF}} = -1.5$ MHz) is applied to produce the double-dressed qubit at 2.5 MHz. Throughout the transitions, the ESR linewidth was noticeably reduced.

The Ramsey interferometry measurement, displayed in the lower graphs of Fig. 2(b)-(d), demonstrates more clearly the effective suppression of inhomogeneous dephasing. The continuous driving protects the qubit from the noisy environment, extending the inhomogeneous dephasing time $T_2^*$. In our measurement, $T_2^*$ of the dressed qubits are $3.34 \pm 0.13$ μs for the single-dressed and $7.23 \pm 1.18$ μs for the double-dressed, which are about 4.5 times and 9.6 times longer, respectively, than that of the bare qubit, $0.75 \pm 0.03$ μs. Note that the oscillations in the Ramsey data are due to the non-zero detuning, except for the fast oscillations shown in Fig. 2(d). The origin of these fast oscillations is unclear, but we suspect they result from spurious beatings between the RF pulse frequency (or the dressed qubit frequency) and the detuning caused by imperfect dressing conditions. See Supplemental Material [43] for more details.

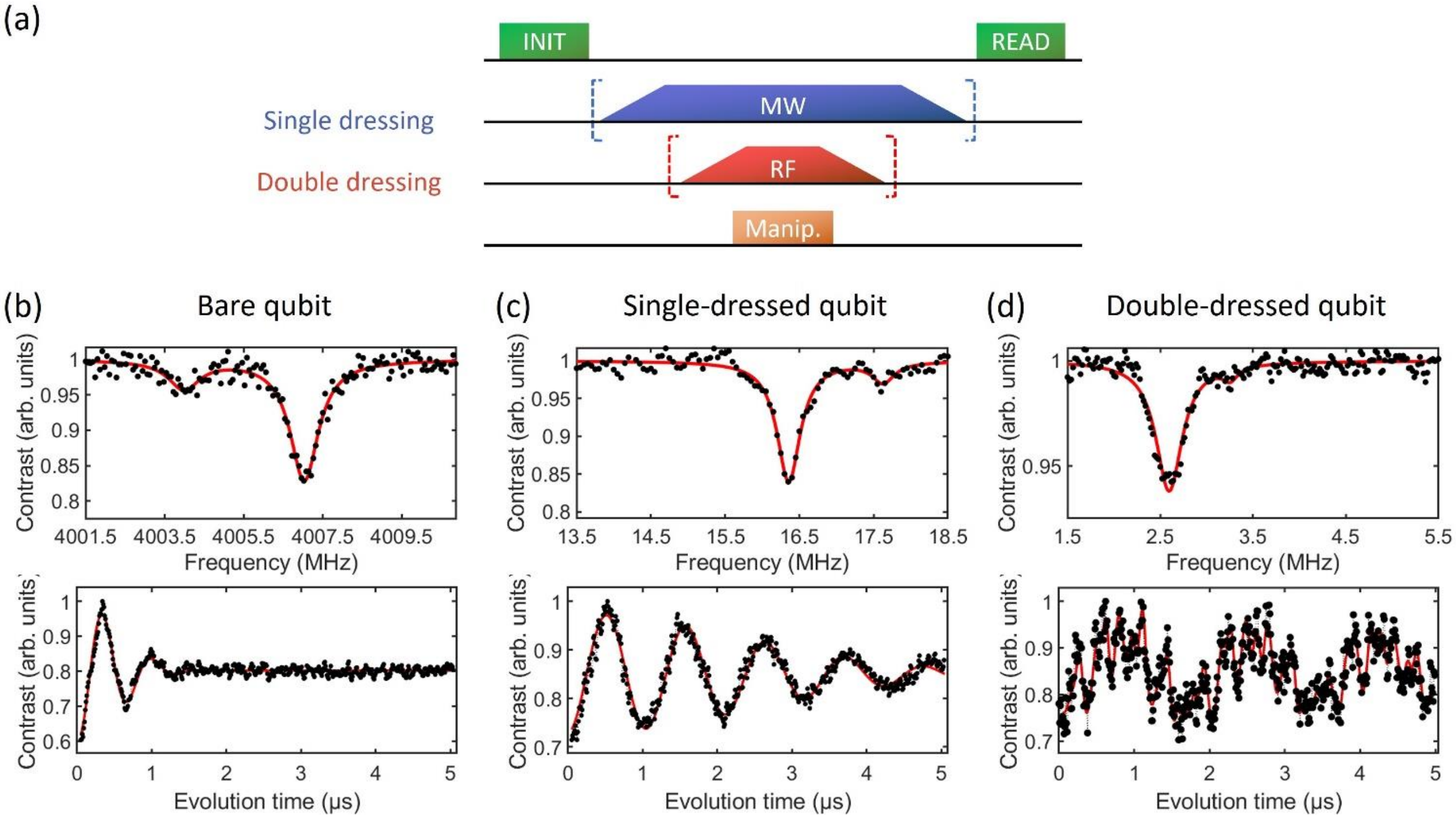

FIG 2. Coherence protection with dressed states (a) Schematics of the measurement sequence used in the double dressing method. From top to bottom: green laser pulses for initialization and readout of the NV center, blue (red) continuous drive for single (double) dressing of qubits with microwave (rf) field, and orange pulses or continuous drive for additional qubit manipulation or external AC field. The driving field amplitudes are gradually adjusted during the adiabatic transition. (b)-(d) ESR (upper panels) and Ramsey (lower panels) measurement results for bare, single-, and double-dressed qubits at frequencies of 4,007 MHz, 16.5 MHz, and 2.5 MHz, respectively. The inhomogeneous dephasing time $T_2^*$ obtained from the fit results of the Ramsey data are $0.75 \pm 0.03$ μs, $3.34 \pm 0.13$ μs, and $7.23 \pm 1.18$ μs respectively. The slow oscillations correspond to the detunings used in the Ramsey measurements. The additional fast oscillations in (d) are discussed in the main text and supplemental materials [43].

## IV. DRESSED STATE SENSING OF AC MAGNETIC FIELD

In the following, we test AC field sensing methods based on dressed state qubits. The method was first introduced in Ref. [36], but we extend it to the double-dressed state. We first spin-lock the qubit along the $x$ axis of the Bloch sphere using a continuous driving of microwave field with an amplitude $\Omega_{\mathrm{MW}}$. When we apply a coherent RF field along the $z$ axis at $\omega_{\mathrm{RF}}$ that matches the dressed qubit frequency, the spin undergoes a population inversion between $|+\rangle = \cos\left(\frac{\theta}{2}\right)|1\rangle + \sin(\frac{\theta}{2})|0\rangle$ and $|-\rangle = -\sin\left(\frac{\theta}{2}\right)|1\rangle + \cos(\frac{\theta}{2})|0\rangle$, resulting in a Rabi oscillation of the single-dressed qubit.

Figure 3(b) illustrates examples of the Rabi measurement at a constant $\omega_{\mathrm{RF}}$ but with different amplitudes $\Omega_{\mathrm{RF}}$. We also perform spectroscopy measurement by sweeping the qubit frequency, as seen in Fig. 3(c). For instance, we plot the Rabi result at a fixed evolution time of $\tau = 4$ μs as a function of $\Omega_{\mathrm{MW}}$, i.e., the single-dressed qubit frequency. Peaks occur only when the qubit frequency matches with $\omega_{\mathrm{RF}}$ and their height is proportional to the Rabi frequency, $\Omega_{\mathrm{RF}}$.

We can do the same measurement with the double-dressed qubit. The spin aligns along the $z$ axis after two consecutive drives of microwave, $\Omega_{\mathrm{MW}}$, and RF field, $\Omega_{\mathrm{RF}}$. We then apply a third coherent field along the $x$ axis, whose frequency, $\omega_{\mathrm{M}}$, corresponding to the double-dressed qubit frequency. Figures 3(d) and (e) are the double-dressed versions of the Rabi and spectroscopy measurements. The Rabi frequency determines the amplitude of the third test field, $\Omega_{\mathrm{M}}$, which shows as a peak in the spectroscopy data.

The combination of coherence protection and AC field sensing, as demonstrated in Figs. 2 and 3, implies that enhanced AC sensing is possible using the double dressing method. The combined effect is especially pronounced at low frequency below 1 MHz. As seen in Figs. 3(c) and (e), the noise floor at the frequency range is elevated owing to the hyperfine interaction with nearby $^{13}$C nuclear spins, which causes spectral broadening and low frequency noise.

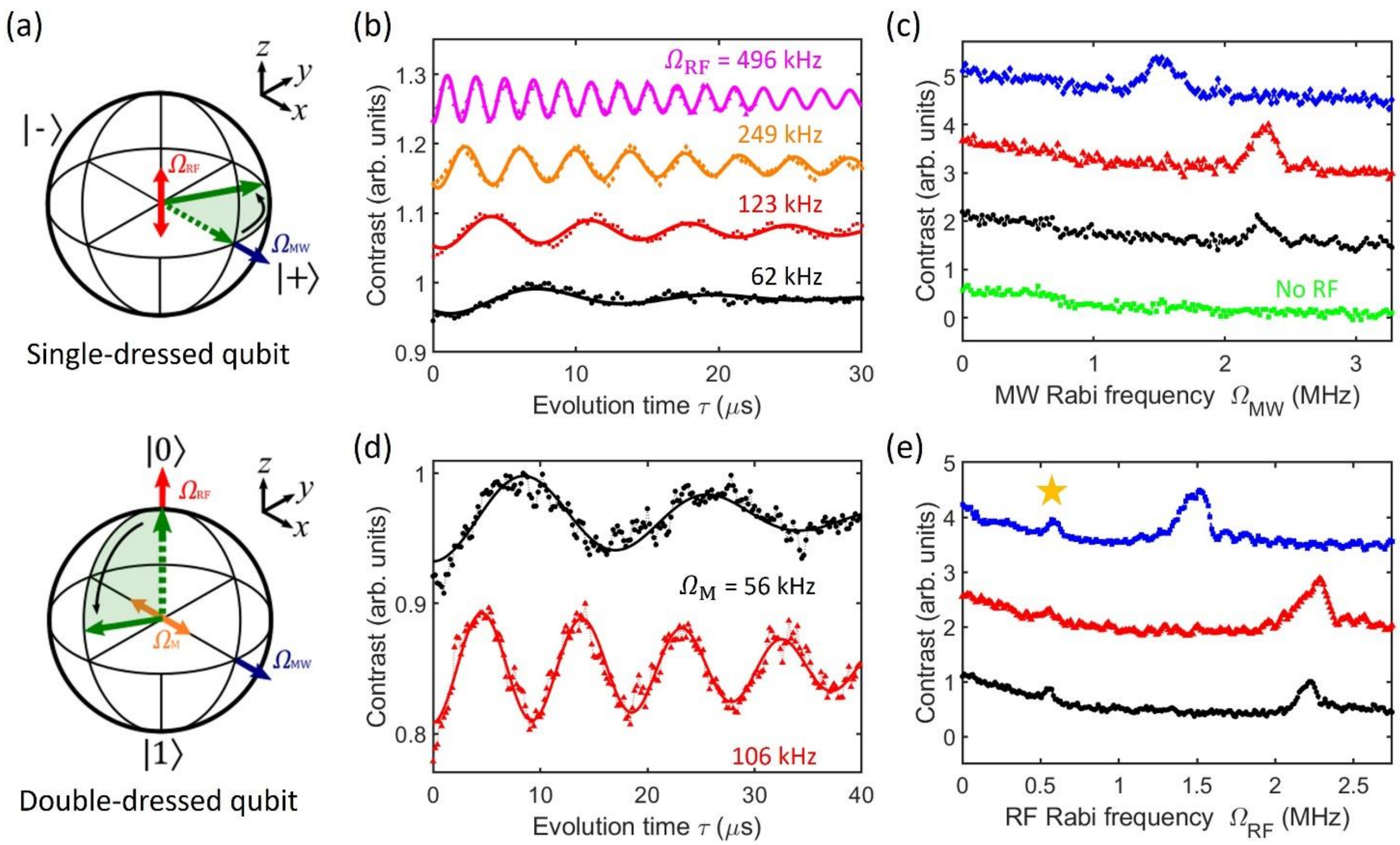

FIG 3. Sensing AC field with dressed states (a) Illustrations of the dressed qubit evolution in the Bloch sphere. (b) and (d) Single- and double-dressed versions of the Rabi oscillation. For single (double)-dressed qubit: a coherent rf (manipulation) field with $\sigma_z$ ($\sigma_x$) component drives the transition between $|+\rangle$ ($|0\rangle$) and $|-\rangle$ ($|1\rangle$), and the oscillation frequency is proportional to $\Omega_{RF}$ ($\Omega_M$). (c) and (e) Noise spectrum measured by the single- and double-dressed qubits. The qubit contrast data in (b) and (d) at fixed evolution times of $\tau = 4$ μs and $\tau = 5$ μs, respectively, are plotted as a function of the qubit frequency, which are tuned by adjusting $\Omega_{MW}$ (for single-dressed) or $\Omega_{RF}$ (for double-dressed). AC signal appears as a peak only when the qubit frequency matches with the signal frequency, and the peak height is roughly proportional to the magnitude of the Rabi frequency in (b) and (d). (c) The applied signal frequencies are 1.50 MHz, 2.23 MHz, 2.23 MHz, with corresponding amplitudes of 4.4 μT, 4.4 μT, 2.2 μT, respectively, listed from top to bottom. (e) The applied signal frequencies are 1.50 MHz, 2.23 MHz, 2.23 MHz, with corresponding amplitudes of 10.4 μT, 10.4 μT, 5.5 μT, respectively, listed from top to bottom. $\delta_{MW}$=3 MHz and $\Omega_{MW}$ = 7.7 MHz are used. See the supplemental material [43] for more information. The small peak in (e) marked by a star is discussed in Fig. 4. All plots in (b)-(e) are y-offset for clarity.

Because of the large noise floor, a high magnetic field is typically required in NMR experiments to push the Larmor frequency of target nuclear spins over 1 MHz, allowing the signal to show against a relatively flat and low noise background.

Utilizing the double dressing method, however, we can effectively lower the sub-MHz noise floor by suppressing the hyperfine interaction and spectral broadening, allowing us to probe target NMR signals like the marked peak around 0.5 MHz in Fig. 3(e). This suggests an alternative method for NMR spectroscopy, even at low magnetic fields. Note that in our proof-of-principle experiment, we applied a magnetic field of ~ 400 G to hyperpolarize the nitrogen nuclear spin. However, this does not affect the core working principle of our method, which remains effective even at lower magnetic fields.

## V. SUPPRESSION OF $^{13}$C NUCLEAR SPIN BATH

Figure 4(a) and 4(b) depict the underlying mechanism of the spin bath suppression at low-frequency regime with a simple toy model calculation. For better visualization and simplicity, we consider six $^{13}$C nuclear spins whose hyperfine couplings (i.e., $A_{\parallel}$ and $A_{\perp}$) are randomly chosen from 0.1 MHz to 1.5 MHz. The corresponding peaks are broadly spread, from 0.5 MHz to 1.2 MHz, at an external field of ~ 405 G. In the calculated spectrum, we also include $^{1}$H nuclear

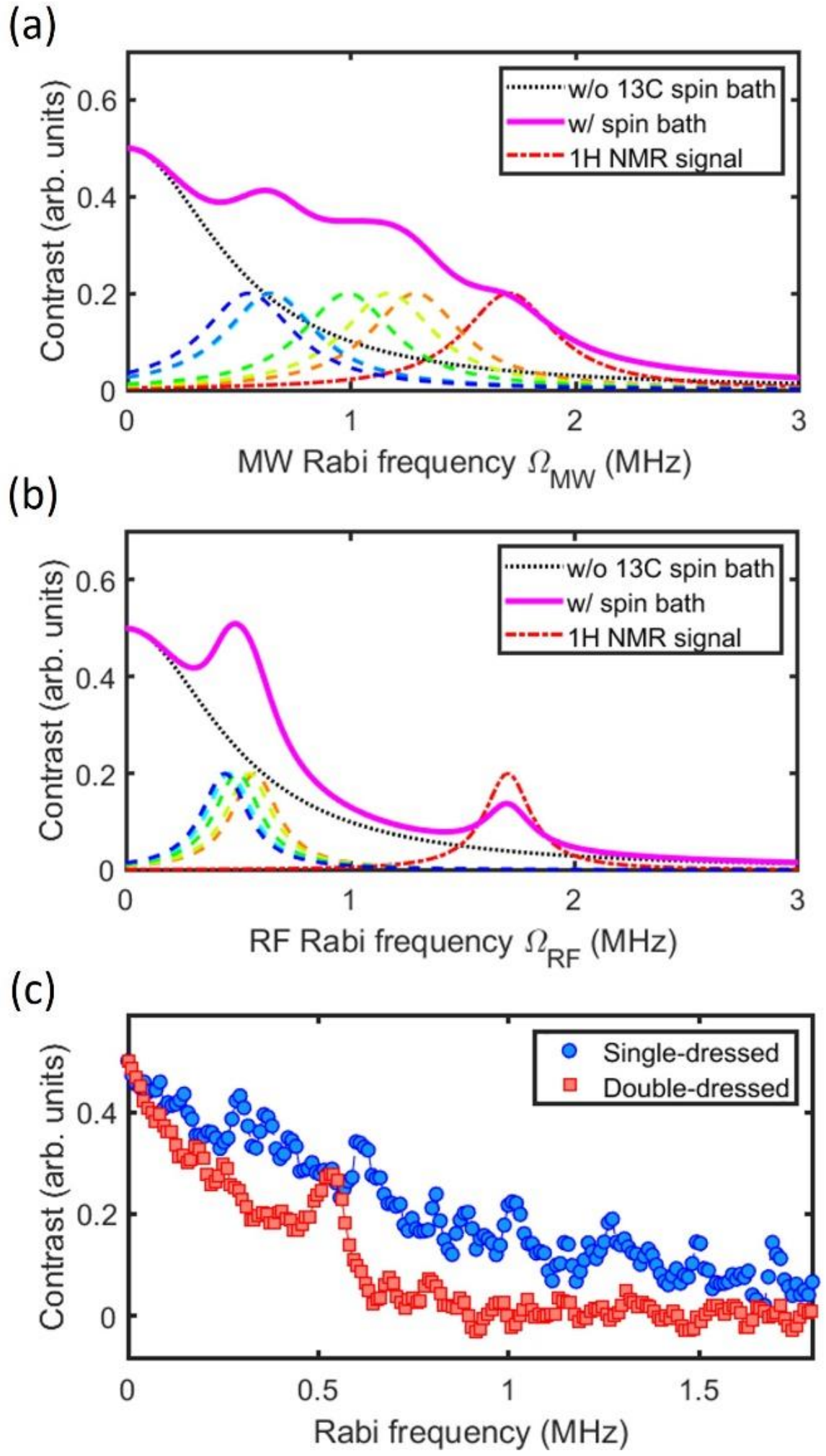

FIG 4. Sub-MHz AC magnetometry based on double-dressed qubit (a, b) Calculated noise spectrum at low frequencies for the (a) single-dressed and (b) double-dressed qubits. In this calculation, we consider six $^{13}$C nuclear spins with hyperfine couplings ranging from 0.1 MHz to 1.5 MHz at an external field of 405 G. We also include $^{1}$H nuclear spins as an example of target NMR spins. Black dotted lines denote low frequency noise other than the spin bath noise. The calculation shows that the widely distributed $^{13}$C peaks shift toward a single peak corresponding to the $^{13}$C Larmor frequency due to the suppression of the hyperfine interaction in the double-dressed qubit. In contrast to the spin bath, the $^{1}$H peak barely moves, suggesting novel NMR sensing of weakly coupled nuclear spins with enhanced SNR. (c) Noise spectrum measured by the single- and double-dressed qubits. Multiple noise spikes in the single-dressed spectrum disappear in the double-dressed spectrum, leaving only a single prominent peak at the expected $^{13}$C Larmor frequency at 405 G.

spins at 1.7 MHz as an example of the weakly coupled NMR targets.

As discussed in Eqs. (2) and (3), the hyperfine couplings are suppressed by a factor of $\sim\frac{\delta_{\mathrm{MW}}}{\omega_{\mathrm{SD}}}$ between the single- and double-dressed qubits. Furthermore, the amount of suppression also depends on the hyperfine coupling itself, which varies with different nuclear spins. Therefore, nuclear spins with stronger hyperfine coupling undergo a larger peak shift in the noise spectrum. In the double-dressed noise spectrum, the spread peaks of six $^{13}$C nuclear spins shift and converge to ~ 0.5 MHz, corresponding to the $^{13}$C Larmor frequency at ~ 405 G. As a result, the sub-MHz noise floor that is predominant by $^{13}$C spin bath is reduced and shifted toward lower frequencies. The $^{1}$H NMR peak, on the other hand, barely shifts due to the relatively much weaker coupling strength.

In Fig. 4 (c), we repeated the measurement from Fig. 3(e) with an increased fixed time of $\tau$ = 10 μs. The multiple spikes in the single-dressed spectrum, which are presumably associated with the $^{13}$C spin bath, are shifted and reduced in the double-dressed spectrum, leaving a single dominant peak at ~ 0.5 MHz, close to the expected location of the $^{13}$C Larmor frequency as discussed in Fig. 4(a-b). Note that $^{1}$H NMR signal from the water layer on the diamond surface was not observed in this measurement, mainly due to the relatively large depth (~ 20 nm) of the NV centers used in this paper. Applying the double-dressed technique to external NMR samples will be a topic for future research.

As an alternative to $^{1}$H signals, we performed additional test measurements using both coherent and stochastic external AC magnetic fields at 0.8 MHz with amplitudes of 3.17 μT. Using the double-dressed protocol, we were able to detect the test signals with an enhanced SNR of up to 17 (see Supplemental Material [43] for more details).

## VI. SUPPRESSION OF LOW FREQUENCY NOISE

In addition to the spin bath, other noise sources can exist in the low-frequency regime, such as magnetic fluctuations from RF sources and external environment. In this section, we test whether such noise can also be suppressed using the dressing methods. This was done by applying stochastic RF noise to the NV center through an RF source, with the noise level varied by adjusting the power gain of the RF amplifier. Power spectral densities of the varied RF noise levels are provided in the Supplemental Material [43].

Additionally, a coherent AC field at 0.6 MHz is applied as a test tone to evaluate its detection against the RF noise. To highlight the performance of the proposed method, we compare the cCDD method (double-dressed state sensing protocol) with the CDD (single-dressed state sensing protocol) and a conventional PDD method, such as XY16. Note that the comparison is made only in the presence of low-frequency noise.

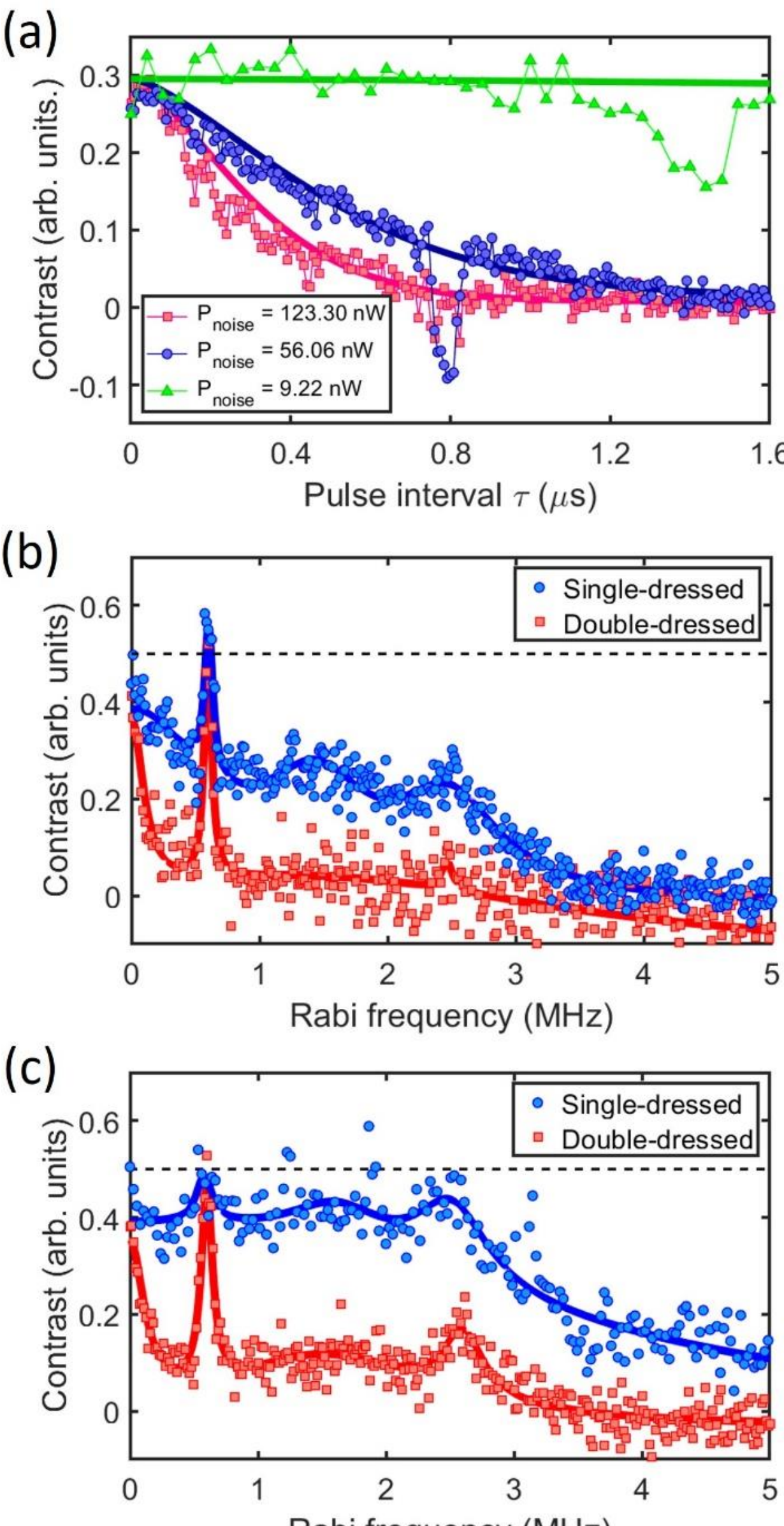

FIG 5. Comparison of PDD, CDD and cCDD as sub-MHz AC magnetometry protocols. (a) XY16 experiments under different RF noise levels. The pulse sequence is $\mathrm{XYXYYXYX}\overline{\mathrm{XYXYYXYX}}$, where X represents in-phase, Y represents quadrature phase, and the bar denotes the $-$ sign. The total evolution time is $N\tau$, where $N$ is the number of $\pi$ pulses and $\tau$ is the pulse interval. Solid lines are fit and serve as visual guidelines for coherence decay. Without the RF amplifier ($P_{\text{noise}} = 9.22$ nW), the $T_2$ coherence time is $274.57 \pm 8.78$ µs and the dip around $\tau$ = 1.4 µs results from weakly coupled $^{13}$C spins. $P_{\text{noise}}$ denotes the RF noise power (see Supplemental Material [43]). When external RF noise is introduced, coherence times decrease to $9.38 \pm 0.33$ µs under moderate RF noise at ($P_{\text{noise}} = 56.06$ nW) and further to $5.70 \pm 0.11$ µs under large noise (at $P_{\text{noise}} = 123.30$ nW). A coherent 0.6 MHz AC field is applied as a test signal. Under moderate noise, the coherence is large enough to detect the signal at $\tau$ = 0.78 µs. However, under large noise, the coherence is insufficient to detect the signal. (b) and (c) Measurements of single- and double-dressed state under RF noise at (b) $P_{\text{noise}} = 56.06$ nW and (c) $P_{\text{noise}} = 123.30$ nW. A 0.6 MHz coherent AC field is applied. The dashed lines indicate the coherence limit, where the coherence is completely lost. The low-frequency noise floor is significantly suppressed, especially in the double-dressed measurements, and the 0.6 MHz signal is detected with enhanced visibility.

Figure 5(a) shows XY16 measurements under different RF noise levels. To probe the 0.6 MHz signal, the optimal pulse interval ($\tau$) for XY16 is 0.78 µs, after considering finite pulse effects [48], resulting in a total sensing time of about 13 µs. However, the NV's coherence time ($T_2$ for PDD and $T_{1\rho}$) decreases as RF noise increases, and reduces from $274.57 \pm 8.78$ µs without the RF amplifier ($P_{\text{noise}} = 9.22$ nW) to $9.38 \pm 0.33$ µs under moderate noise ($P_{\text{noise}} = 56.06$ nW) and $5.70 \pm 0.11$ µs under large noise ($P_{\text{noise}} = 123.30$ nW). As a result, the signal at 0.6 MHz is barely visible under large noise, though it remains detectable under moderate noise. Note that the 0.6 MHz test tone was not applied in the case of $P_{\text{noise}} = 9.22$ nW, as the RF amplifier was turned off.

In contrast, for the CDD and cCDD measurements, the signal is clearly visible for both moderate and large RF noise levels, particularly with the cCDD method. This improvement is mainly due to the suppression of low-frequency noise in these protocols.

For a more quantitative analysis, we introduce the metric "visibility", defined as the normalized maximal contrast after decoherence (a detailed calculation of visibility is available in Appendix E). For the coherence-limited measurement, the estimated visibility under RF noise with $P_{\text{noise}} = 56.06$ nW (RF noise with $P_{\text{noise}} = 123.30$ nW) is 0.25 (0.07) for XY16, 0.48 (0.21) for CDD, and 0.92 (0.87) for cCDD, respectively. Figure 5 clearly demonstrates the cCDD method's ability to suppress noise floors and enhance visibility, even under significant low-frequency noise.

## VII. CONCLUSIONS

In conclusion, we introduce the cCDD method and demonstrate its effectiveness suppressing sub-MHz noise, including the hyperfine interaction between the NV center and the $^{13}$C nuclear spin bath, as well as external noise. With the help of continuous microwave and RF drives, we adiabatically transfer the NV center into the single- and double-dressed qubits. From the Ramsey experiments, we observe an extension of the effective dephasing time ($T_2^*$), confirming the suppression of low-frequency noise, including qubit-bath interactions, energy level fluctuations, and other noise sources. Additionally, we show a substantial reduction in low-frequency noise, below ~ 1 MHz due to suppressed spectral broadening and reduced noise associated with both qubit-bath interactions and external noise. These improvements enable the detection of sub-MHz signals with enhanced SNR and visibility. This work provides a novel method for low-field NMR and low-frequency AC magnetometry based on diamond NV centers.

## DATA AVAILABILITY

The data used in this study are available in Ref. [56].

## ACKNOWLEDGMENTS

This work is supported by the National Research Foundation of Korea (NRF) grant (NRF-2019M3E4A1080232, NRF-2022M3K4A1094777), Information & communication Technology Planning & Evaluation (IITP) grant (No. RS-2023-00230717, IITP-2024-2020-0-01606) funded by the Korea government (MSIT).

## APPENDIX A: EXPERIMENTAL SETUP

We use a home-built confocal scanning fluorescence microscopy to address the NV centers. The NV centers used in experiments in Figs. 2 - 4 are fabricated in an electronic grade diamond with natural abundance of 1.1% $^{13}$C. For the comparison measurements in Fig. 5, we use isotopically purified diamond of 0.001 % $^{13}$C concentration [7] and the applied magnetic field is 328 G. Using a $^{13}$C purified diamond, while spin noise from other defects or nitrogen nuclear spins still exists, classical noise becomes the primary noise source. We apply external magnetic field with a permanent magnet attached to a 3-axis piezoelectric scanner. See Supplemental Material [43] for more details.

## APPENDIX B: SENSING PROTOCOL

The dressed state sensing protocol is accomplished by matching the Hartmann-Hahn condition. The cross relaxation between matched energy levels allows us to determine the frequency and amplitude of the interaction. Unlike conventional pulsed dynamical decoupling sensing protocols, this method is a spurious-free protocol, making it a promising tool for analyzing complex NMR signals [46].

## APPENDIX C: NOISE MODEL

In Fig. 5, we fit the noise floor of the dressed state relaxometry using the power spectrum density model from [39]. We calculate SNR or visibility from the fitted noise level.

## APPENDIX D: BANDWIDTH OF THE N PULSES PDD

Increasing the number of pulses ($N$) increases the coherence time ($T_2$) by $N^m$ times, where $0 < m < 1$ [50]. The coherence time of total spin evolution time ($N\tau$) is extended while the upper bound for the pulse interval $\tau$ is decreased, resulting in a narrower bandwidth. Thus, the sensitivity and bandwidth are often traded off. With larger $N$, the sensitivity increases but only for a high frequency band, which makes it challenging to achieve high sensitivity in low frequency magnetometry using the PDD method.

## APPENDIX E: CALCULATION OF THE VISIBILITY

Visibility is the normalized maximal contrast after the decoherence over evolution time $t$. To calculate coherence decay in the PDD measurements, we use two sequential measurements with different projection operators: $+\frac{\pi}{2}_{\mathrm{X}}$ pulse and $-\frac{\pi}{2}_{\mathrm{X}}$ pulse. We refer to these measurement as $S^+(t)$ and $S^-(t)$, respectively.

The normalized coherence $C^+(t)$ and $C^-(t)$ can be calculated using the reference measurement $I_{|0\rangle}$, which is the fluorescence intensity of $|0\rangle$;

$$C^+(t) = \frac{S^+(t)}{I_{|0\rangle}}, \text{ and } C^-(t) = \frac{S^-(t)}{I_{|0\rangle}}. \quad (4)$$

Subtracting the two values yields a noiseless signal,

$$S(t) = C^+(t) - C^-(t) \quad (5)$$

, with a range of around $-0.3$ to 0.3 and it converges to 0; a classical state.

$S(t)$ is normalized as

$$\tilde{S}(t) = \frac{S(t) - S(t)_{\mathrm{min}}}{2S(t)_{\mathrm{max}}} \quad (6)$$

, or equivalently,

$$\tilde{S}(t) = \left(1 + \frac{(S^+(t) - I_{|1\rangle}) - (S^-(t) - I_{|1\rangle})}{I_{|0\rangle} - I_{|1\rangle}}\right)/2 \quad (7)$$

, where $S(t)_{\mathrm{max}} = -S(t)_{\mathrm{min}} = (I_{|0\rangle} - I_{|1\rangle})/I_{|0\rangle}$.

Now we calculate the maximum possible contrast, i.e., visibility, which is twice of the reminiscent coherence at time $t$,

$$V(t) = 2\left(\tilde{S}(t) - \frac{1}{2}\right) = \frac{S^+(t) - S^-(t)}{I_{|0\rangle} - I_{|1\rangle}}. \quad (8)$$

Eq. (8) is used to calculate the visibility in the PDD experiments.

Note that, for the CDD and cCDD measurements, we only measure $S^-(t)$ and normalize it with the Rabi reference measurement, which has already been characterized from the dressing parameters. It is because $S^{+}(t)$ and $S^{-}(t)$ are equivalent measurements and, therefore, $S^{+}(t)+S^{-}(t)=I_{|0\rangle}+I_{|1\rangle}$ = constant.

The visibilities for these measurements are obtained from

$$V(t)=\frac{\left(I_{|0\rangle}+I_{|1\rangle}\right)-2S^{-}(t)}{I_{|0\rangle}-I_{|1\rangle}} = 2\left(\frac{1}{2}-\frac{S^{-}(t)-I_{|1\rangle}}{I_{|0\rangle}-I_{|1\rangle}}\right). \quad (9)$$

# Supplemental material

## Suppression of spin bath for sub-MHz AC magnetometry based on double-dressed spin qubit in diamond

Kihwan Kim[1], Yisoo Na[1], Jungbae Yoon[1], Dongkwon Lee[1], Hee Seong Kang[2], Chul-Ho Lee[3], Chulki Kim[4], and Donghun Lee[1]

[1] Department of Physics, Korea University, Seoul, 02841, Republic of Korea
[2] KU-KIST Graduate School of Converging Science and Technology, Korea University, Seoul, 02841, Republic of Korea
[3]Department of Electrical and Computer Engineering, Seoul National University, Seoul, 08826, Republic of Korea
[4]Quantum Information Center, Korea Institute of Science and Technology (KIST), Seoul, 02792, Republic of Korea

## Supplementary Note 1: Diamond sample

Commercial electronic-grade single crystal diamonds are used in this paper (Element 6). The samples were implanted with nitrogen ions at 7° tilt angle, $1 \times 10^9/\mathrm{cm}^2$, and 15 keV (Innovion) and then Ar annealed at 850 ℃ for 2 hours to create NV centers. After annealing, the sample was treated with tri-acids for 2 hours and oxygen annealed at 465 ℃ for 4 hours to make the NV centers' charge state stable.

For the comparison experiments between XY16, CDD and cCDD, an isotopically purified diamond is used, which is the same diamond as used in [1].

## Supplementary Note 2: Confocal microscope

A home-built confocal scanning microscope is used to image and readout the diamond NV centers (Fig. S1). 532 nm laser is controlled by an acousto-optic modulator (AOM, IntraAction ATM-200C1) to generate laser pulse. The 532 nm excitation beam and the NV's fluorescence signal of ~ 630 – 800 nm are separated by a dichroic mirror. The fluorescence signal is spatially filtered with a confocal pinhole of 50 μm diameter and then focused to the single photon detector, avalanche photon detector (APD, Excelitas SPCM-AQRH-14-FC). A band pass filter and a notch filter block reminiscent green laser going in to the APD. The external DC magnetic field is applied by a permanent magnet maneuverable with a 3-axis translation stage.

## Supplementary Note 3: Microwave and rf control electronics

Two Keysight M8190 arbitrary waveform generator (AWG) modules (total 4 channels) with a Keysight M8192 synchronization module generate microwave (~ GHz) and rf (~ MHz) continuous waves or pulses that are used to control the spin qubit. The control fields are further amplified by microwave or rf amplifiers that are combined by a bias tee to be delivered to the NV centers with a coplanar wave guide (CPW) fabricated on a glass substrate or a bonded gold wire.

## Supplementary Note 4: Adiabatic passage

We adiabatically map the dressed qubit state to the bare qubit state so that the coherent control and readout of the qubit state is possible. Previous works have discussed about adiabatic passage and suggested methods to speed up the process [2–5]. For instance, adding a counter-diabatic driving to a resonant driving will speed up the adiabatic process and enhance the process fidelity [3]. However, the technique requires precise design of the phase and amplitude of the control pulse at each time. On the other hand, dressing with a non-zero finite detuning can be an alternative and a relatively easy way to realize fast adiabatic passage. In this paper, we adopted the latter method. As discussed in Fig. 2(a), we linearly ramp the amplitude of detuned driving fields and gradually transfer the qubit to the target dressed state. As far as the energy along the principal axis is maintained larger than the increasing transverse energy, we can ramp the amplitude faster than the resonant dressing case without losing the adiabaticity.

As the detuning and the ramping time are competing to each other to maintain the adiabaticity, we need to find their optimal parameters. We first do simulation of the qubit evolution with various transfer conditions using an open-source Quantum Toolbox in Python, QuTiP[6, 7].

Figure S2 and S3 are examples of the simulation. Starting from $|0\rangle$, the qubit evolves to $|+\rangle$ and returns back to $|0\rangle$. As seen in Fig. S2(a), when the detuning is 1.5 MHz and the Rabi frequency is 15 MHz, 2 μs ramping time is not slow enough to maintain the adiabaticity and the final state fidelity (how much back to the initial state, $|0\rangle$) is 0.919 after total evolution time of 10 μs including two 2 μs ramping time and 6 μs locking time. This can be improved by two different methods. First method is simply increasing the ramping time. Figure S2(b) is the same process as Fig. S2(a) but with ten times longer ramping time, i.e. 20 μs. The adiabaticity recovers and the final state fidelity becomes 0.999 after total evolution time of 50 μs (two 20 μs ramping time and 10 μs locking time). Another solution is introducing larger detuning, for example 4 MHz, while keeping the ramping time as 2 μs. As shown in Fig. S3(a), high state fidelity comparable to the first method is realized. In this paper, we use the latter method to have the ramping time as short as possible. Note that there exist other adiabatic methods such as using the counter-diabatic driving [3] as shown in Fig. S3(b).

Figure S4 compares experimental results with two different adiabatic conditions, i.e. resonant ramping and ramping with non-zero detuning. For the former process, the ramping time is not long enough and results in oscillations during the ramping and imperfect recovery to the initial state. On the other hand, the latter process does not involve oscillations and the qubit returns to the initial state with higher fidelity. Note that due to the non-zero detuning, however, the dressed principal axis is not perfectly orthogonal to the bare basis axis but tilted with a certain angle of $\theta$ which is determined by the Rabi frequency $\Omega$ and the detuning $\delta$ through the relation of $\tan(\theta) = \frac{\Omega}{\delta}$. Also note that diabatic evolution can still exist even if we try to adiabatically dress the qubit, especially when the qubit is doubly dressed. The spurious signal in Fig. 2(d) can be explained by this diabatic evolution. The diabatically evolved populations should undergo incoherent dynamics (incoherent with the adiabatic passage but still coherent with additional driving) so that there can be additional fast oscillations in the Ramsey interferometry. Figure S5 shows the fast Fourier transformation (FFT) of the double-dressed Ramsey data used in Fig. 2(d). The Ramsey data were fitted by including five oscillation frequencies from the FFT: 0.6 MHz, 2.4 MHz, 3.0 MHz, 3.6 MHz, and 6.0 MHz. We suspect these frequencies are harmonics of the applied driving (3 MHz) and the detuning (0.6 MHz) used in the double-dressed measurement, arising due to imperfect adiabatic conditions, possibly resulting from an insufficient ramping time. A thorough understanding of the effects of diabatic evolution is beyond the scope of this paper and needs to be addressed in future investigations.

## Supplementary Note 5: Pulse sequences for the double dressing method

We briefly explain our qubit control pulse sequence used in Fig. 2(a). In Fig. S6, we visualize the pulse sequence that the AWG makes to realize the double-dressed Ramsey experiment. The frequencies of the applied driving fields in Fig. S6 are not real values used in the experiments but are chosen to enhance the visibility of the waves. The laser pulse controls the AOM for initialization and readout of the NV center. The sampling trigger goes to the data acquisition system (DAQ) which collect the photon count data from the APD.

## Supplementary Note 6: Dressed ESR experiments

Throughout the experiments, an external DC field of ~ 400 G is applied, and the nitrogen nuclear spin is polarized. The ESR experiments in Fig. 2(a) shows that the $^{15}$N nuclear spin is polarized to $m_{\mathrm{I}}$ = +1/2. Therefore, we ignore the $^{15}$N hyperfine split doublet and treat the electron spin states of $m_{\mathrm{s}}$ = +1 and $m_{\mathrm{s}}$ = 0 as the effective two-level system. Our bare qubit frequency is set to be 4007.4 MHz which is associated with $m_{\mathrm{I}}$ = +1/2 sub-state. When we apply 4 MHz blue-detuned microwave driving with 16 MHz Rabi frequency, the single-dressed qubit energy becomes 16.5 MHz. A similar process occurs for the double-dressed qubit whose frequency becomes 2.5 MHz driven by the 1.5 MHz red-detuned rf driving with 2.0 MHz Rabi frequency.

## Supplementary Note 7: ESR linewidth narrowing due to suppression of hyperfine coupling

Here we do a simple toy model calculation for the ESR linewidth narrowing shown in Fig. 2(b)-(d). We assume a hyperfine split doublet consisting of the ESR curve as shown in Fig. S7. In brief, dressing the qubit reduces the hyperfine coupling and thus results in the linewidth narrowing. Given the original electron resonance frequencies of $f_1$, $f_2$, Rabi frequency $\Omega$, detuning $\delta$ and the hyperfine splitting of $A$, the dressed resonance frequencies become $\sqrt{\delta^2 + \Omega^2}$ and $\sqrt{(\delta + A)^2 + \Omega^2}$ (for simplicity, the non-secular terms are ignored). If we assume $\Omega \gg \delta > A$, which is an easily fulfilled condition in pure diamond samples as the hyperfine interactions are usually less than 1 MHz [8] while one can easily fulfill the Rabi frequency over 10 MHz, we can write the new energy splitting of $\sqrt{(\delta + A)^2 + \Omega^2} - \sqrt{\delta^2 + \Omega^2}$ as,

$$\sqrt{(\delta + A)^2 + \Omega^2} - \sqrt{\delta^2 + \Omega^2} = \Omega\left(1 + \left(\frac{\delta + A}{\Omega}\right)^2\right)^{\frac{1}{2}} - \Omega\left(1 + \left(\frac{\delta}{\Omega}\right)^2\right)^{\frac{1}{2}}$$

$$\approx \Omega\left(1 + \frac{1}{2}\left(\frac{\delta + A}{\Omega}\right)^2\right) - \Omega\left(1 + \frac{1}{2}\left(\frac{\delta}{\Omega}\right)^2\right) = \frac{A^2 + 2A\delta}{2\Omega} \qquad \text{(S1)}$$

, which is about $\frac{\delta}{\Omega}$ times smaller than the original splitting of $A$. Thus, the energy splitting can be easily reduced by more than an order of magnitude in the most cases depending on the magnitude of detuning, which is also easy to be controlled. Figure S7 shows an example plot of the energy splitting versus the Rabi frequency when $A$ = 0.3 MHz and a relatively large detuning of $\delta = 3$ MHz.

## Supplementary Note 8: Derivation of the dressed Hamiltonian : resonant cases

Here we derive Hamiltonians for the single- and double-dressed states. For the singly dressing case, the bare qubit Hamiltonian is written with one driving field,

$$H \approx \frac{\omega_q}{2}\sigma_z + \sigma_z \sum_{\vec{r}}\left[\frac{A_{||,\vec{r}}}{2} I_{z,\vec{r}} + \frac{A_{\perp,\vec{r}}}{2} I_{x,\vec{r}}\right] + \sum_{\vec{r}} \gamma_{n,\vec{r}} B I_{z,\vec{r}} + \epsilon\sigma_z + \Omega_{MW}\cos(\omega_{MW} t)\,\sigma_x. \text{ (S2)}$$

In terms of the environmental noise, we consider two terms; hyperfine interaction with the nuclear spin bath and qubit energy fluctuation along the *z* axis, $\epsilon$. This time we ignore the transverse noise as it contributes little as discussed in the main text. As a dressing field, we consider a continuous microwave drive whose frequency is $\omega_{MW}$ and amplitude is $\Omega_{MW}$. Without loss of generality, we consider the field is oscillating along the *x* axis. We move to the rotating frame via a unitary transformation $\hat{U}$, so that the time dependency to be removed,

$$\hat{U} = e^{i\frac{\omega_{MW}}{2} t \sigma_z}. \text{ (S3)}$$

Then after the rotating wave approximation (RWA), the Hamiltonian becomes,

$$H_{\mathrm{SD}} \approx \frac{\delta_{\mathrm{MW}}}{2}\sigma_{\mathrm{z}} + \frac{\Omega_{\mathrm{MW}}}{2}\sigma_{\mathrm{x}} + \sum_{\vec{r}}\left[\frac{A_{||,\vec{r}}}{2}\sigma_{\mathrm{z}} I_{\mathrm{z},\vec{r}} + \frac{A_{\perp,\vec{r}}}{2}\sigma_{\mathrm{z}} I_{\mathrm{x},\vec{r}}\right] + \sum_{\vec{r}}\gamma_{\mathrm{n},\vec{r}} B I_{\mathrm{z},\vec{r}} + \epsilon\sigma_{\mathrm{z}} \quad \text{(S4)}$$

, where $\delta_{\mathrm{MW}} = \omega_{\mathrm{q}} - \omega_{\mathrm{MW}}$. The noise terms maintain their original forms as they commute with $\sigma_{\mathrm{z}}$. This can be further simplified by defining a variable $\theta_{\mathrm{SD}}$ such that $\tan(\theta_{\mathrm{SD}}) = \frac{\Omega_{\mathrm{MW}}}{\delta_{\mathrm{MW}}}$.

With the continuous driving, the initial bare qubit basis $\{|0\rangle, |1\rangle\}$ along the *z* axis is transformed to the new basis $\{|+\rangle, |-\rangle\}$, which is now tilted to some degrees from the *z* axis. This is called the dressed basis,

$$\begin{bmatrix}|+\rangle \\ |-\rangle\end{bmatrix} = \begin{bmatrix}\cos(\frac{\theta_{\mathrm{SD}}}{2}) & \sin(\frac{\theta_{\mathrm{SD}}}{2}) \\ \sin(\frac{\theta_{\mathrm{SD}}}{2}) & -\cos(\frac{\theta_{\mathrm{SD}}}{2})\end{bmatrix}\begin{bmatrix}|1\rangle \\ |0\rangle\end{bmatrix}. \quad \text{(S5)}$$

If we consider $\delta_{MW} = 0$ for simplicity, the Hamiltonian in eq. S4 becomes

$$H_{\mathrm{SD}} \approx \frac{\Omega_{\mathrm{MW}}}{2}\sigma_{\mathrm{x}} + \sum_{\vec{r}}\left[\frac{A_{||,\vec{r}}}{2}\sigma_{\mathrm{z}} I_{\mathrm{z},\vec{r}} + \frac{A_{\perp,\vec{r}}}{2}\sigma_{\mathrm{z}} I_{\mathrm{x},\vec{r}}\right] + \sum_{\vec{r}}\gamma_{\mathrm{n},\vec{r}} B I_{\mathrm{z},\vec{r}} + \epsilon\sigma_{\mathrm{z}}. \quad \text{(S6)}$$

Now we move to the double-dressed state. Among several known ways to doubly dress a spin qubit system [9~13], we choose the double dressing method using a secondary rf driving along the *z* axis. The bare qubit Hamiltonian is written as,

$$H \approx \frac{\omega_{\mathrm{q}}}{2}\sigma_{\mathrm{z}} + \sigma_{\mathrm{z}}\sum_{\vec{r}}\left[\frac{A_{||,\vec{r}}}{2} I_{\mathrm{z},\vec{r}} + \frac{A_{\perp,\vec{r}}}{2} I_{\mathrm{x},\vec{r}}\right] + \sum_{\vec{r}}\gamma_{\mathrm{n},\vec{r}} B I_{\mathrm{z},\vec{r}} + \epsilon\sigma_{\mathrm{z}} + \Omega_{\mathrm{MW}}\cos(\omega_{\mathrm{MW}}\mathrm{t})\,\sigma_{\mathrm{x}} +$$
$$\Omega_{\mathrm{RF}}\cos(\omega_{\mathrm{RF}}\mathrm{t})\,\sigma_{\mathrm{z}}. \quad \text{(S7)}$$

Compared to eq. S2, we now have two driving terms, i.e. microwave and rf. Again for simplicity we drive the microwave field resonantly ($\delta_{\mathrm{MW}} \approx 0$). Then the single-dressed Hamiltonian becomes

$$H_{\mathrm{SD}} \approx \frac{\Omega_{\mathrm{MW}}}{2}\sigma_{\mathrm{x}} + \sum_{\vec{r}}\left[\frac{A_{||,\vec{r}}}{2}\sigma_{\mathrm{z}} I_{\mathrm{z},\vec{r}} + \frac{A_{\perp,\vec{r}}}{2}\sigma_{\mathrm{z}} I_{\mathrm{x},\vec{r}}\right] + \sum_{\vec{r}}\gamma_{\mathrm{n},\vec{r}} B I_{\mathrm{z},\vec{r}} + \epsilon\sigma_{\mathrm{z}} + \Omega_{\mathrm{RF}}\, cos(\omega_{\mathrm{RF}} t)\,\sigma_{\mathrm{z}}. \quad \text{(S8)}$$

If we change the frame to be on the new basis,

$$\widetilde{H}_{\mathrm{SD}} \approx \frac{\Omega_{\mathrm{MW}}}{2}\tilde{\sigma}_{\mathrm{z}} + \Omega_{\mathrm{RF}}\, cos(\omega_{\mathrm{RF}} t)\,\tilde{\sigma}_{\mathrm{x}} + \sum_{\vec{r}}\left[\frac{A_{||,\vec{r}}}{2}\tilde{\sigma}_{\mathrm{x}} I_{\mathrm{z},\vec{r}} + \frac{A_{\perp,\vec{r}}}{2}\tilde{\sigma}_{\mathrm{x}} I_{\mathrm{x},\vec{r}}\right] + \sum_{\vec{r}}\gamma_{\mathrm{n},\vec{r}} B I_{\mathrm{z},\vec{r}} + \epsilon\tilde{\sigma}_{\mathrm{x}} \quad \text{(S9)}$$

, where $\tilde{\sigma}_{\mathrm{z}} = \sigma_{\mathrm{x}}$ and $\tilde{\sigma}_{\mathrm{x}} = \sigma_{\mathrm{z}}$. The two leading terms in eq. S9 looks just the same as a bare qubit with a driving field. We now move to the new rotating frame, which is the double-dressed frame.

$$H_{\mathrm{DD}} \approx \frac{\delta_{\mathrm{RF}}}{2}\tilde{\sigma}_{\mathrm{z}} + \frac{\Omega_{\mathrm{RF}}}{2}\tilde{\sigma}_{\mathrm{x}} + \sum_{\vec{\mathrm{r}}} e^{i\frac{\omega_{\mathrm{RF}}}{2}t\tilde{\sigma}_{\mathrm{z}}}\tilde{\sigma}_{\mathrm{x}} e^{-i\frac{\omega_{\mathrm{RF}}}{2}t\tilde{\sigma}_{\mathrm{z}}}\left[\frac{A_{\|,\vec{\mathrm{r}}}}{2}I_{\mathrm{z},\vec{\mathrm{r}}} + \frac{A_{\perp,\vec{\mathrm{r}}}}{2}I_{\mathrm{x},\vec{\mathrm{r}}}\right] + \epsilon e^{i\frac{\omega_{\mathrm{RF}}}{2}t\tilde{\sigma}_{\mathrm{z}}}\tilde{\sigma}_{\mathrm{x}} e^{-i\frac{\omega_{\mathrm{RF}}}{2}t\tilde{\sigma}_{\mathrm{z}}} + \sum_{\vec{\mathrm{r}}}\left[\gamma_{\mathrm{n},\vec{\mathrm{r}}} B I_{\mathrm{z},\vec{\mathrm{r}}}\right] \quad \text{(S10)}$$

, where $\delta_{\mathrm{RF}} = \Omega_{\mathrm{MW}} - \omega_{\mathrm{RF}}$. If we assume the second drive is also on resonance, $\delta_{\mathrm{RF}} = 0$ and change the frame to be on the new double-dressed basis, the Hamiltonian becomes,

$$H_{\mathrm{DD}} \approx \frac{\Omega_{\mathrm{RF}}}{2}\sigma_{\mathrm{z}} + \sum_{\vec{\mathrm{r}}} e^{i\frac{\omega_{\mathrm{RF}}}{2}t\tilde{\sigma}_{\mathrm{z}}}\tilde{\sigma}_{\mathrm{x}} e^{-i\frac{\omega_{\mathrm{RF}}}{2}t\tilde{\sigma}_{\mathrm{z}}}\left[\frac{A_{\|,\vec{\mathrm{r}}}}{2}I_{\mathrm{z},\vec{\mathrm{r}}} + \frac{A_{\perp,\vec{\mathrm{r}}}}{2}I_{\mathrm{x},\vec{\mathrm{r}}}\right] + \epsilon e^{i\frac{\omega_{\mathrm{RF}}}{2}t\tilde{\sigma}_{\mathrm{z}}}\tilde{\sigma}_{\mathrm{x}} e^{-i\frac{\omega_{\mathrm{RF}}}{2}t\tilde{\sigma}_{\mathrm{z}}} + \sum_{\vec{\mathrm{r}}} \gamma_{\mathrm{n},\vec{\mathrm{r}}} B I_{\mathrm{z},\vec{\mathrm{r}}} \quad \text{(S11)}$$

, whose quantization axis is back to the *z* axis as same as the bare qubit. The nuclear spin Zeeman terms, or the NMR signals, survive throughout the transformations as they commute with the unitary transforms we apply. The hyperfine interaction and the energy fluctuation terms are further removed by RWA and the Hamiltonian finally becomes,

$$H_{\mathrm{DD}} \approx \frac{\Omega_{\mathrm{RF}}}{2}\sigma_{\mathrm{z}} + \sum_{\vec{\mathrm{r}}} \gamma_{\mathrm{n},\vec{\mathrm{r}}} B I_{\mathrm{z},\vec{\mathrm{r}}}. \quad \text{(S12)}$$

In this regime, the double dressing once more decouples the qubit from the noisy spin bath and the environmental fluctuation. Note that through the hierarchy of the dressing procedure, the ratio of $\omega_{\mathrm{RF}}$ to $\Omega_{\mathrm{MW}}$ for the double-dressed qubit is smaller compared to the ratio of $\omega_{\mathrm{MW}}$ to $\omega_q$ for the bare qubit. Although the approximation used to derive the Hamiltonians is still valid, there may exist non-zero higher order noise terms in the double-dressed Hamiltonian making the spin qubit still vulnerable to the noisy environment.

## Supplementary Note 9: Hamiltonian calculation with spin 1 system

In the main text, we derive the qubit Hamiltonian based on the two-level system. Here we show the Hamiltonian calculation based on the S = 1 spin system and consider an effective two-level system. The Hamiltonian for the single dressing case is written as, (we treat $\hbar$ = 1)

$$H = DS_{\mathrm{z}}^2 + \gamma_{\mathrm{e}} B S_{\mathrm{z}} + \Omega\cos(\omega t)\, S_{\mathrm{x}} \quad \text{(S13)}$$

, where *D* is the zero-field splitting, *B* is the applied magnetic field, $\gamma_{\mathrm{e}}$ is the gyromagnetic ratio of the electron spin and *S*=$(S_{\mathrm{x}}, S_{\mathrm{y}}, S_{\mathrm{z}})$ is the spin 1 operator. The Hamiltonian can be expressed in the matrix form based on the basis $\{|{+1}\rangle, |0\rangle, |{-1}\rangle\}$ as,

$$\begin{bmatrix} D+\gamma_e B & \frac{1}{\sqrt{2}}\,\Omega\cos(\omega t) & 0 \\ \frac{1}{\sqrt{2}}\,\Omega\cos(\omega t) & 0 & \frac{1}{\sqrt{2}}\,\Omega\cos(\omega t) \\ 0 & \frac{1}{\sqrt{2}}\,\Omega\cos(\omega t) & D-\gamma_e B \end{bmatrix}. \text{ (S14)}$$

We move to the rotating frame via a unitary transform,

$$\hat{U} = e^{i\omega t S_z^2}. \text{ (S15)}$$

Note that we use $S_z^2$ in the exponent of the operator as the large zero field splitting is dominant
in the Hamiltonian. Then the Hamiltonian becomes,

$$\begin{bmatrix} D+\gamma_e B-\omega & \frac{\Omega}{2\sqrt{2}} & 0 \\ \frac{\Omega}{2\sqrt{2}} & 0 & \frac{\Omega}{2\sqrt{2}} \\ 0 & \frac{\Omega}{2\sqrt{2}} & D-\gamma_e B-\omega \end{bmatrix} \text{ (S16)}$$

, and we define $\delta = D - \gamma_e B - \omega$. If we tune the frequency of the driving field to the $|0\rangle \leftrightarrow$
$|-1\rangle$ transition, so that $|+1\rangle$ state is far detuned by several GHz, we can ignore it and treat
the system as an effective two-level system. For simplicity, we shift the energy level by $-\delta/2$
so that the reduced two-level Hamiltonian becomes,

$$\frac{1}{2}\begin{bmatrix} -\delta & \frac{\Omega}{\sqrt{2}} \\ \frac{\Omega}{\sqrt{2}} & \delta \end{bmatrix}. \text{ (S17)}$$

Once we define $\tan(\theta) = \left(\frac{\Omega}{\sqrt{2}}\right)/\delta$, the single dressed basis becomes,

$$\begin{bmatrix} |+\rangle \\ |-\rangle \end{bmatrix} = \begin{bmatrix} \sin\left(\frac{\theta}{2}\right) & \cos\left(\frac{\theta}{2}\right) \\ \cos\left(\frac{\theta}{2}\right) & -\sin\left(\frac{\theta}{2}\right) \end{bmatrix} \begin{bmatrix} |0\rangle \\ |-1\rangle \end{bmatrix} \text{ (S18)}$$

, which is the same results as eq. S5

**Supplementary Note 10: Hyperfine splitting on the dressed basis**

This section explains the spin 1 Hamiltonian with hyperfine interactions. Consider spin 1/2 $^{13}$C
and the secular approximation, the Hamiltonian reads,

$$H = DS_z^2 + \gamma_e B S_z + A_{||} S_z I_z + A_{\perp} S_z I_x + \gamma_n B I_z \text{ (S19)}$$

, where $A_{||}$ and $A_{\perp}$ are the parallel and perpendicular hyperfine couplings, and $\gamma_{\mathrm{n}}$ is the
nuclear Larmor frequency. As $m_{\mathrm{s}} = +1$ resonance is a few GHz away from the others, we
will focus on only the $m_{\mathrm{s}} = 0$ and $m_{\mathrm{s}} = -1$ sub-spin systems as an effective qubit. When
the driving field $\Omega \cos(\omega t) S_{\mathrm{x}}$ is added, the Hamiltonian becomes,

$$H = \begin{bmatrix} \frac{\gamma_{\mathrm{n}} B}{2} & 0 & \frac{\Omega \cos(\omega t)}{\sqrt{2}} & 0 \\ 0 & -\frac{\gamma_{\mathrm{n}} B}{2} & 0 & \frac{\Omega \cos(\omega t)}{\sqrt{2}} \\ \frac{\Omega \cos(\omega t)}{\sqrt{2}} & 0 & D - \gamma_{\mathrm{e}} B + \frac{\gamma_{\mathrm{n}} B - A_{||}}{2} & -\frac{A_{\perp}}{2} \\ 0 & \frac{\Omega \cos(\omega t)}{\sqrt{2}} & -\frac{A_{\perp}}{2} & D - \gamma_{\mathrm{e}} B + \frac{A_{||} - \gamma_{\mathrm{n}} B}{2} \end{bmatrix}. \quad \text{(S20)}$$

We move to the interaction picture via a unitary transform $\hat{U} = e^{i\omega t S_{\mathrm{z}}^2} \otimes \hat{I}$ and, after RWA,
the matrix becomes,

$$H_{\mathrm{RWA}} = \begin{bmatrix} \frac{\gamma_{\mathrm{n}} B}{2} & 0 & \frac{\Omega}{2\sqrt{2}} & 0 \\ 0 & -\frac{\gamma_{\mathrm{n}} B}{2} & 0 & \frac{\Omega}{2\sqrt{2}} \\ \frac{\Omega}{2\sqrt{2}} & 0 & D - \gamma_{\mathrm{e}} B - \omega + \frac{\gamma_{\mathrm{n}} B - A_{||}}{2} & -\frac{A_{\perp}}{2} \\ 0 & \frac{\Omega}{2\sqrt{2}} & -\frac{A_{\perp}}{2} & D - \gamma_{\mathrm{e}} B - \omega + \frac{A_{||} - \gamma_{\mathrm{n}} B}{2} \end{bmatrix}. \quad \text{(S21)}$$

Once we define $\delta = D - \gamma_{\mathrm{e}} B - \omega$,

$$H_{\mathrm{RWA}} = \begin{bmatrix} \frac{\gamma_{\mathrm{n}} B}{2} & 0 & \frac{\Omega}{2\sqrt{2}} & 0 \\ 0 & -\frac{\gamma_{\mathrm{n}} B}{2} & 0 & \frac{\Omega}{2\sqrt{2}} \\ \frac{\Omega}{2\sqrt{2}} & 0 & \delta + \frac{\gamma_{\mathrm{n}} B}{2} - \frac{A_{||}}{2} & -\frac{A_{\perp}}{2} \\ 0 & \frac{\Omega}{2\sqrt{2}} & -\frac{A_{\perp}}{2} & \delta - \frac{\gamma_{\mathrm{n}} B}{2} + \frac{A_{||}}{2} \end{bmatrix}. \quad \text{(S22)}$$

Numerical calculation of eq. S22 is shown in Fig. S9 which plots the relative ratio of the
hyperfine splitting between the single-dressed qubit and the bare spin qubit as a function of $A_{||}$
and $A_{\perp}$. For most of the parameter range, the ratio is larger than one meaning the hyperfine
splitting becomes smaller after the dressing.

In the main text, we assumed the “large detuning” regime for easier calculation of eq. S22 and
derived the suppression factor of $\frac{\delta_{\mathrm{MW}}}{\sqrt{\delta_{\mathrm{MW}}^2 + \Omega_{\mathrm{MW}}^2}} \approx \frac{\delta_{\mathrm{MW}}}{\Omega_{\mathrm{MW}}}$. This factor is valid for most of cases even including small detuning or resonance. Figure S10 shows numerical calculation of two different detuning regimes; large detuning and small detuning (nearly resonant). The red region in the contour plot of Fig. S10(a) indeed confirms that the ratio agrees well with the suppression factor of $\frac{\delta_{\mathrm{MW}}}{\Omega_{\mathrm{MW}}}$. Unless we are close to the minimum at $A_{||} \sim 0.867$ MHz, where $A_{||}$ cancels the nuclear Zeeman interaction, the factor is valid for most of the parameter space, especially near the zero point where most of the nuclear spin bath are responsible for. For the small detuning case in Fig. S10(b), the difference between the numerically calculation and the approximation ($\frac{\delta_{\mathrm{MW}}}{\Omega_{\mathrm{MW}}}$) is about 10-20 % in the orange area. The discrepancy becomes larger for larger values of $A_{||}$ and $A_{\perp}$.

## Supplementary Note 11: Calculations used in Fig. 4(a)

The hyperfine interaction transforms when the electron spin is dressed. When the electron spin is singly dressed, it is locked on the transvers plane and the nuclear spins experience a new effective magnetic field including the hyperfine interaction,

$$\vec{B} - \frac{1}{2}(A_{\perp}, 0, A_{||}) \quad \text{(S23)}$$

, where $\vec{B}$ is the static magnetic field in the lab frame [12]. Thus the new effective Larmor frequency of the nuclear spins becomes,

$$\sqrt{(\gamma_{\mathrm{n}} B - \frac{A_{||}}{2})^2 + (\frac{A_{\perp}}{2})^2}\,. \quad \text{(S24)}$$

The frequencies of the six $^{13}$C nuclear spins in the upper graph of Fig. 4(a) are calculated from eq. S24 with randomly chosen values of $A_{||}$ and $A_{\perp}$ from 0.1 MHz to 1.5 MHz. When the electron spin is doubly dressed as shown in the lower graph of Fig. 4(a), the hyperfine couplings in eq. S24 are reduced by a factor of $\sim \frac{\delta_{\mathrm{MW}}}{\omega_{\mathrm{SD}}}$ and the peaks converge to the bare Larmor frequency of $^{13}$C nuclear spin at the given external field of $\vec{B}$.

## Supplementary Note 12: AC sensing measurement

As discussed in Fig. 4, we test sub-MHz AC magnetometry using both stochastic and coherent signals. We applied both coherent and stochastic fields at 0.8 MHz with amplitudes of 3.17 μT as test signals. For instance, Fig. S12(c) shows that the coherent signal is clearly visible in the double-dressed spectrum as a result of a 12 dB suppression in the noise floor. This measurement enables sub-MHz AC sensing with an enhanced SNR ≈ 17. The pulse sequences used for the measurements are illustrated in Fig. S11. Figure S12 (a) and (b) shows the before and after of introducing randomly phased stochastic AC signal at 0.8 MHz. While the signal is not observed in the single-dressed measurements, it is clearly visible in the double-dressed data with SNR ≈ 8. Note that the small bump around 0.4 MHz may correspond to the $^{13}$C Larmor frequency at an external magnetic field of ~ 377 G.

Now, we present detailed calculations on the strength of the applied AC fields. Our single qubit Hamiltonian is given as below,

$$H_{\mathrm{q}} = \frac{\omega_{\mathrm{q}}}{2}\sigma_{\mathrm{z}}.$$

All of the applied AC fields, i.e., MW driving, RF driving, and the signal, are intended to be on the x-z plane without loss of generality. We consider terms that are resonant throughout the dressing procedure that the MW and RF Hamiltonian become,

$$H_{\mathrm{MW}} = \Omega_{\mathrm{MW,x}}\cos(\omega_{\mathrm{MW}}t)\,\sigma_{\mathrm{x}},$$

$$H_{\mathrm{RF}} = \Omega_{\mathrm{RF,z}}\cos(\omega_{\mathrm{RF}}t)\,\sigma_{\mathrm{z}}.$$

The AC signal (or the manipulation pulse in Fig. 2 and Fig. 3 in the main text) in the double-dressed frame is

$$H_{\mathrm{S}} = \Omega_{S,\mathrm{z}}\cos(\omega_{\mathrm{S}}t)\,\sigma_{\mathrm{z}}.$$

The total Hamiltonian becomes,

$$H = H_{\mathrm{q}} + H_{\mathrm{MW}} + H_{\mathrm{RF}} + H_{\mathrm{S}}$$

When we test the single-dressed magnetometry, $H_{\mathrm{RF}}$ serves as the signal, and therefore no additional $H_{\mathrm{S}}$ is required.

In the single-dressed frame with the MW detuning $\delta_{\mathrm{MW}} \equiv \omega_{\mathrm{q}} - \omega_{\mathrm{MW}}$,

$$H_{\mathrm{SD}} = H_{\mathrm{q,SD}} + H_{\mathrm{RF,SD}} + H_{\mathrm{S,SD}}$$

with,

$$H_{\mathrm{q,SD}} \approx \frac{\omega_{\mathrm{SD}}}{2}\sigma_{\mathrm{z,SD}},$$

$$H_{\mathrm{RF,SD}} = \Omega_{\mathrm{RF,z}}\cos(\omega_{\mathrm{RF}}t)\left[\frac{\delta_{\mathrm{MW}}}{\Omega_{\mathrm{SD}}}\sigma_{\mathrm{z,SD}} + \frac{\Omega_{\mathrm{MW,x}}}{\Omega_{\mathrm{SD}}}\sigma_{\mathrm{x,SD}}\right],$$

$$H_{\mathrm{S,SD}} = \varOmega_{S,\mathrm{z}} \cos(\omega_{\mathrm{S}} t) \left[\frac{\delta_{\mathrm{MW}}}{\varOmega_{\mathrm{SD}}} \sigma_{\mathrm{z,SD}} + \frac{\varOmega_{\mathrm{MW,x}}}{\varOmega_{\mathrm{SD}}} \sigma_{\mathrm{x,SD}}\right],$$

where we transformed the Hamiltonians using the unitary operator $U = e^{i\frac{\omega_{\mathrm{MW}} t}{2}\sigma_{\mathrm{z}}}$ and $\omega_{\mathrm{SD}} = \sqrt{{\delta_{\mathrm{MW}}}^2 + {\varOmega_{\mathrm{MW,x}}}^2}$. First, we consider the single-dressed magnetometry. With RWA, the single-dressed qubit frequency becomes $\omega_{SD}$ and the perpendicular driving term becomes $\varOmega_{\mathrm{RF,z}} \cos(\omega_{\mathrm{RF}} t) \frac{\varOmega_{\mathrm{MW,x}}}{\varOmega_{\mathrm{SD}}} \sigma_{\mathrm{x,SD}}$. On resonant ($\delta_{\mathrm{MW}} = 0$), the original magnitude of the RF driving field along the z axis is preserved. However, for non-zero detuning ($\delta_{\mathrm{MW}} \neq 0$), the RF amplitude is projected to the tilted dressed x-y plane, yielding in a reduction of the amplitude by $\frac{\varOmega_{\mathrm{MW,x}}}{\varOmega_{\mathrm{SD}}}$.

In the double-dressed frame with unitary operator $U = e^{i\frac{\omega_{\mathrm{RF}} t}{2}\sigma_{\mathrm{z,SD}}}$, the Hamiltonian becomes,

$$H_{\mathrm{DD}} = H_{\mathrm{q,DD}} + H_{\mathrm{S,DD}}$$

$$H_{\mathrm{q,DD}} \approx \frac{\omega_{\mathrm{DD}}}{2} \sigma_{\mathrm{z,SD}}$$

$$H_{\mathrm{S,DD}} \approx \varOmega_{S,\mathrm{z}} \frac{\delta_{\mathrm{MW}}}{\varOmega_{\mathrm{SD}}} \cos(\omega_{\mathrm{S}} t) \left[\frac{\delta_{\mathrm{RF}}}{\omega_{\mathrm{DD}}} \sigma_{\mathrm{z,DD}} + \frac{\varOmega_{\mathrm{DD}}}{\omega_{\mathrm{DD}}} \sigma_{\mathrm{x,DD}}\right],$$

where $\delta_{\mathrm{RF}} \equiv \omega_{\mathrm{SD}} - \omega_{\mathrm{RF}}$, $\varOmega_{\mathrm{DD}} \equiv \varOmega_{\mathrm{RF,z}} \frac{\varOmega_{\mathrm{MW,x}}}{\varOmega_{\mathrm{SD}}}$, and $\omega_{\mathrm{DD}} = \sqrt{{\delta_{\mathrm{RF}}}^2 + {\varOmega_{\mathrm{DD}}}^2}$.

The double-dressed magnetometry can be explained by the Rabi model with the double-dressed qubit energy of $\omega_{\mathrm{DD}}$ and the perpendicular driving term of $\varOmega_{S,\mathrm{z}} \frac{\delta_{\mathrm{MW}}}{\varOmega_{\mathrm{SD}}} \cos(\omega_{\mathrm{S}} t) \frac{\varOmega_{\mathrm{DD}}}{\omega_{\mathrm{DD}}} \sigma_{\mathrm{x,DD}}$.

In summary, the signal strength drops by $\frac{\delta_{\mathrm{MW}}}{\varOmega_{\mathrm{SD}}} \frac{\varOmega_{\mathrm{DD}}}{\sqrt{\delta_{\mathrm{RF}}^2 + \varOmega_{\mathrm{DD}}^2}}$ for the double-dressed case. In Fig. 5 of the main text, for example, MW detuning is $\delta_{\mathrm{MW}}$ ~ 0.125 MHz for the single-dressed case which is from an additional $^{13}\mathrm{C}$ nuclear spin with the hyperfine splitting of ~ 0.250 MHz. For the double-dressed magnetometry, $\varOmega_{\mathrm{SD}}$ = 14.56 MHz and $\delta_{\mathrm{MW}}$ = 9 MHz. As the signal frequency is 0.6 MHz, the suppression factors are $\frac{\varOmega_{\mathrm{MW,x}}}{\varOmega_{\mathrm{SD}}} = 0.979$ for the single-dressed case and $\frac{\delta_{\mathrm{MW}}}{\varOmega_{\mathrm{SD}}} \frac{\varOmega_{\mathrm{DD}}}{\sqrt{\delta_{\mathrm{RF}}^2 + \varOmega_{\mathrm{DD}}^2}} = 0.613$ for the double-dressed case. We applied 1.67 times larger signal in the double-dressed experiment to achieve a similar Rabi frequency as in the single-dressed case. Although the signal amplitude decreases, the SNR can be increased, as seen in Fig. 5, because the double-dressing suppresses the noise floor considerably more than the signal. The experiments in Fig. S12 did not reveal any extra $^{13}\mathrm{C}$ nuclear spins within the $T_2^*$ limit. However, if we assume any detuning smaller than the $T_2^*$ limit, the ratio between single-dressed and double-dressed Rabi frequencies becomes roughly 2 and we indeed applied 2.67 times larger signal to reach comparable Rabi frequency in the double-dressed experiment (we used $\Omega_{\mathrm{SD}}$ = 13.3 MHz and $\delta_{\mathrm{MW}}$ = 5 MHz). Again, despite the reduction in signal, there is larger suppression in noise, resulting in increased SNR.

## Supplementary Note 13: XY16 and spin locking coherence time

XY16 and single-dressed spin locking experiments with no RF noise show a coherence time of $274.57 \pm 8.78\ \mu\mathrm{s}$ (~ $16\tau$) as in Fig. S13. The dips correspond to $^{13}$C nuclear spin bath. The single-dressed coherence time, $T_{1,\rho}$, was measured and found to be $346.84 \pm 49.32\ \mu\mathrm{s}$ as in Fig. S14.

## Supplementary Note 14: Characterization of the RF noise

An oscilloscope (Keysight, DSAV134A) is used to characterize the RF noise used in Fig. 5. The RF amplifier specifies with a maximum power gain of 53 dB at 1 MHz. We adjusted the RF noise levels by tuning the amplifier's power gain ($G_{\mathrm{dB}}$). The resulting power spectral densities (PSDs) are shown in Fig. S15. To quantify the RF noise power ($P_{\mathrm{noise}}$), we integrated the noise power spectrum over the range from 0 Hz to 5 MHz. The resulting values of $P_{\mathrm{noise}}$ are 9.22 nW, 56.06 nW, 71.67 nW, 123.30 nW for gain settings of 0, 36.4 dB, 42.4 dB and 46.0 dB, respectively.

References for supplemental material

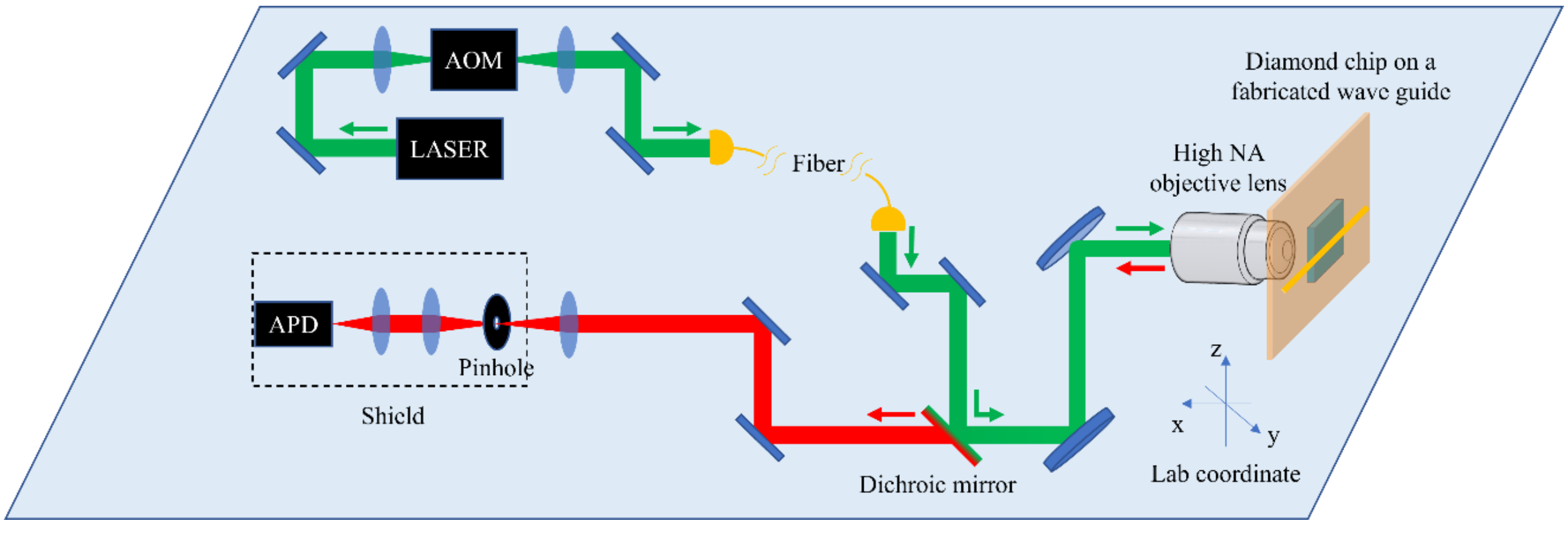

369
370
371 Fig. S1. Schematics of the experiment setup
372
373

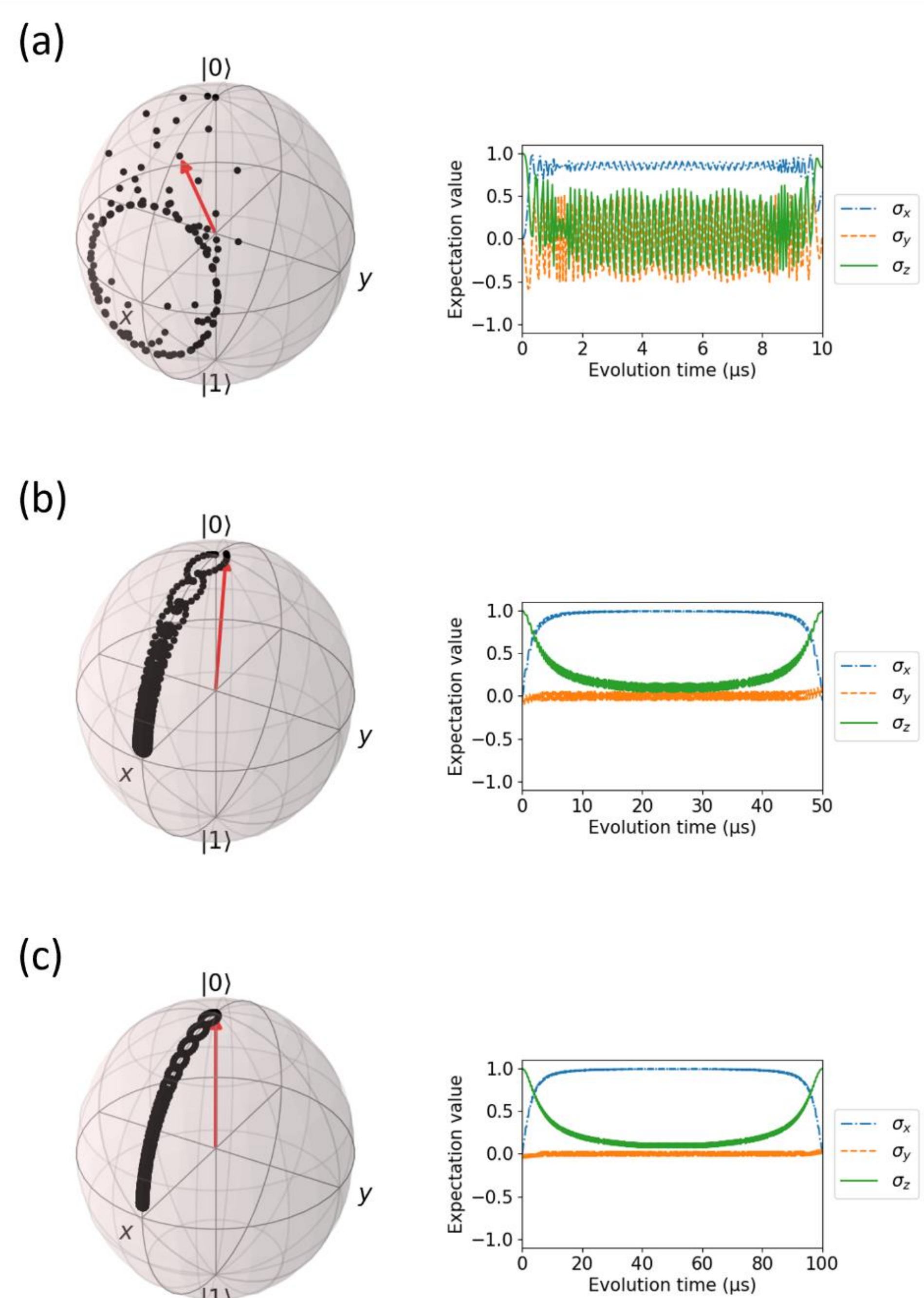

Fig. S2. QuTiP simulation of adiabatic passage. Spin evolves from $|0\rangle$ to $|1\rangle$ with various
ramping time of (a) 2 μs, (b) 20 μs, and (c) 40 μs. Rabi frequency of 15 MHz and detuning
of 1.5 MHz are the same for all the cases. The expectation values of the x, y, and z components
are plotted together with the Bloch sphere representations

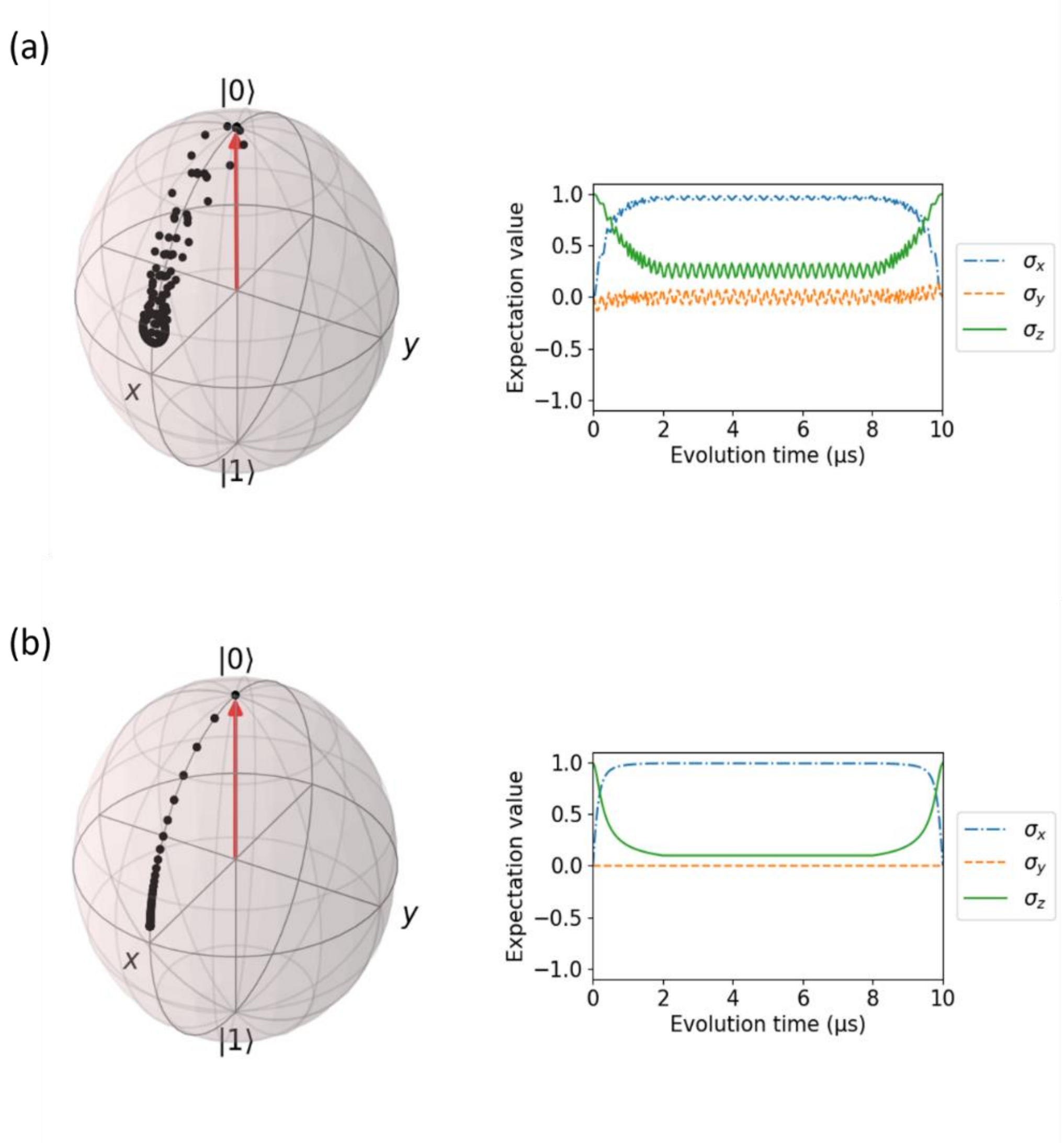

Fig. S3. QuTiP simulation of adiabatic passage. (a) Detuning is increased to 4 MHz compared to Fig. S2(a). Ramping time (2 μs) and Rabi frequency (15 MHz) are the same as Fig. S2(a). (b) Counter-diabatic driving method. Two drives are used with the same parameters but different axes, i.e. x for one drive and y for the other. 2 μs ramping time, 15 MHz Rabi frequency and 1.5 MHz detuning are used.

392
393

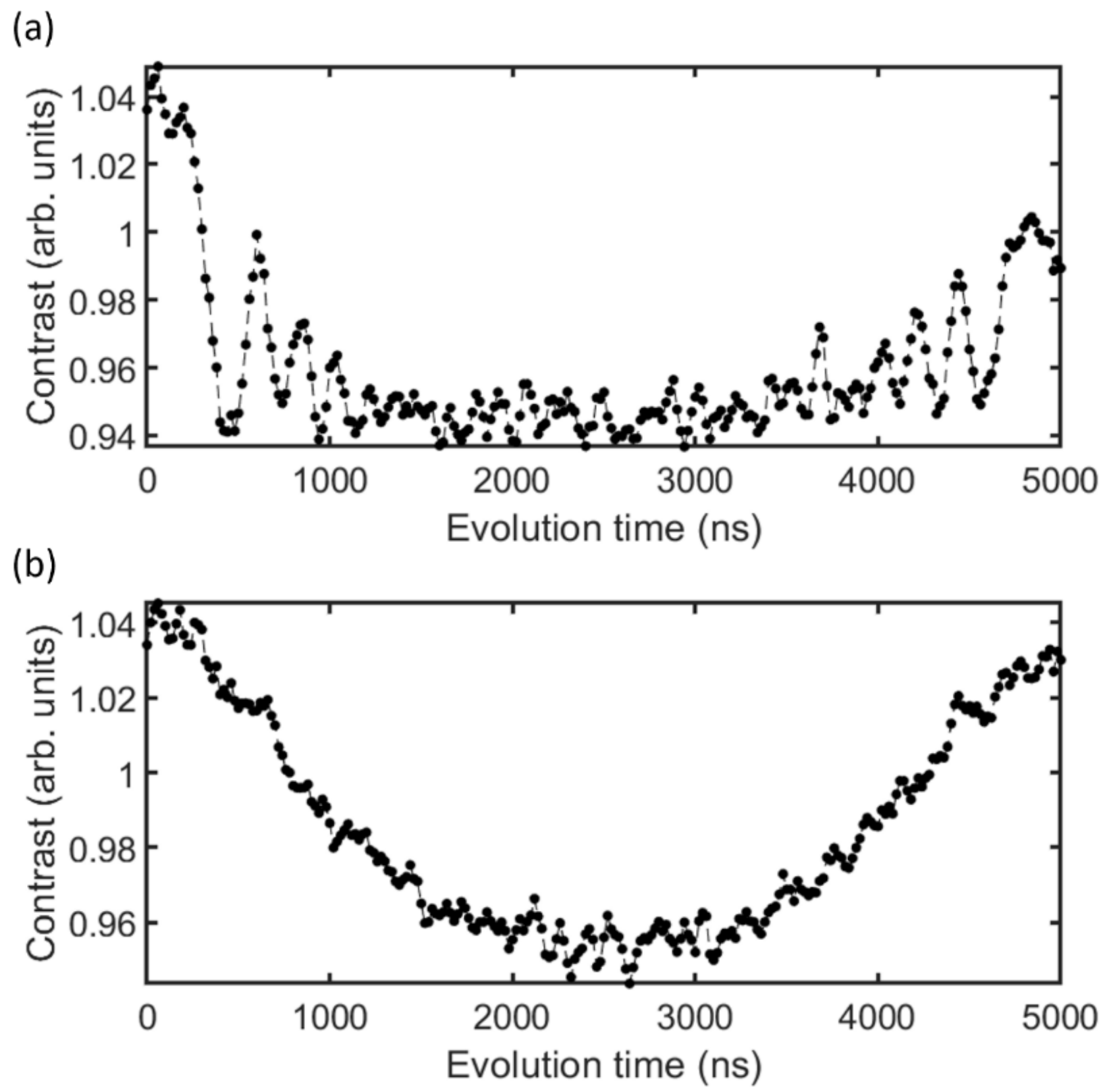

Fig. S4. Testing the adiabatic passage. (a) Resonant ramping, and (b) ramping with 3 MHz
detuning. Rabi frequency is 7.5 MHz for both cases.

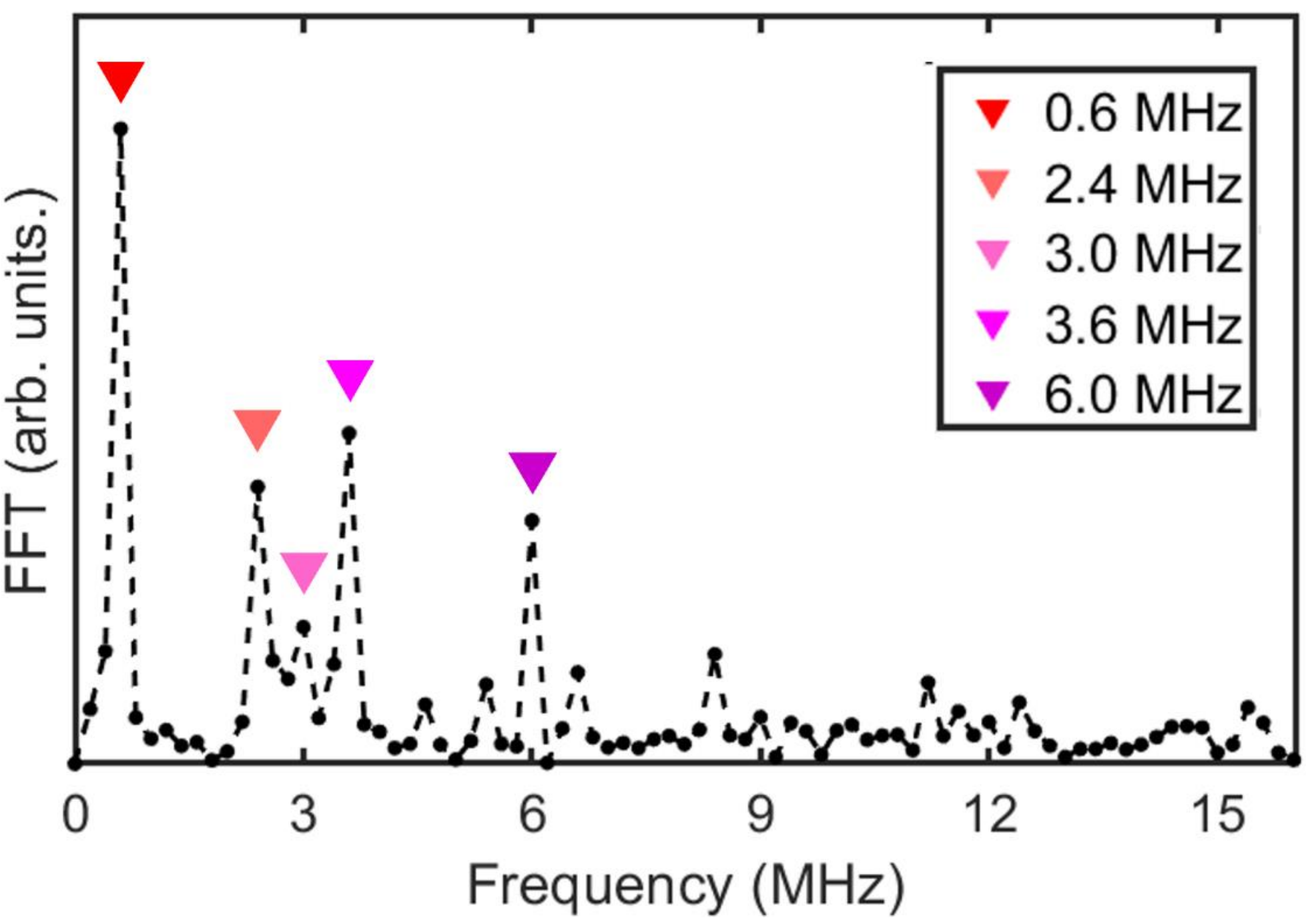

Fig. S5. Fast Fourier transformation (FFT) of the double dressed Ramsey data used in Fig. 2(d).

The five most dominant frequencies were used to fit the Ramsey data.

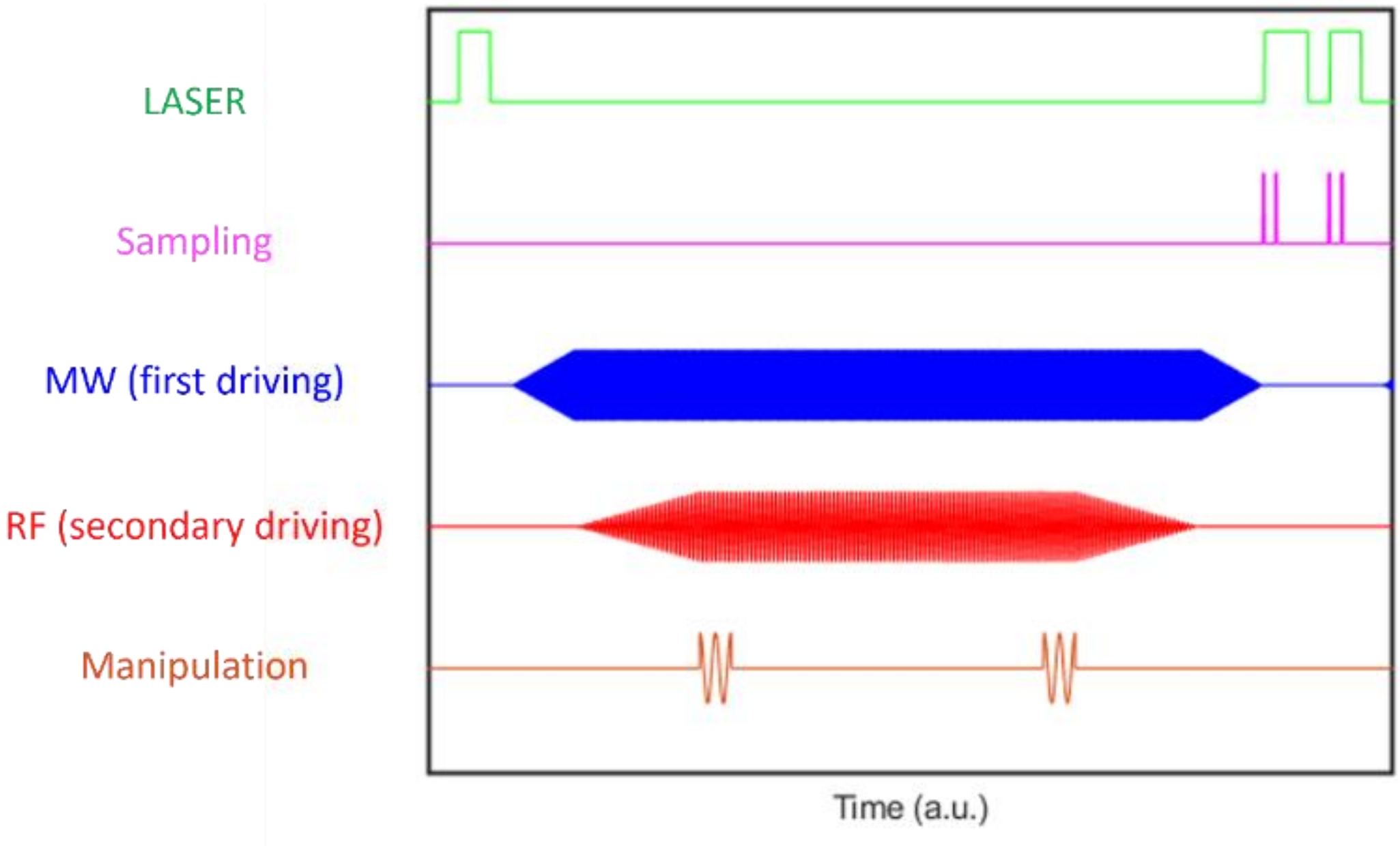

Fig. S6. A pulse sequence used for the double-dressed Ramsey experiment.

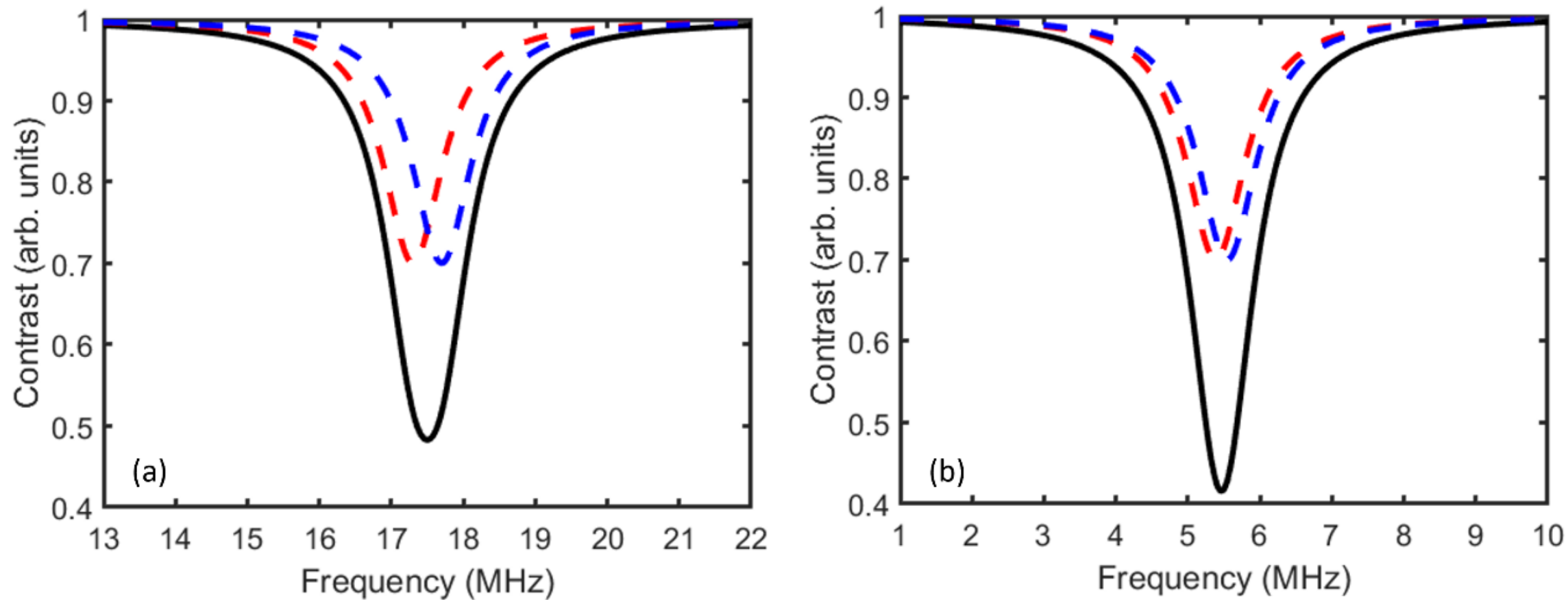

Fig. S7. Dressing reduces ESR linewidth owing to the suppression of hyperfine coupling – a toy model. The two dashed curves correspond to the hyperfine split doublets that form the ESR curve (solid line); (a) and (b) are before and after the dressing field. Upon application of continuous driving (5 MHz Rabi frequency and 2 MHz detuning), the hyperfine splitting is reduced and the ESR linewidth is narrowed by a factor of ~ 2/5. The intrinsic linewidth is assumed to be 0.3 MHz for both cases.

424
425

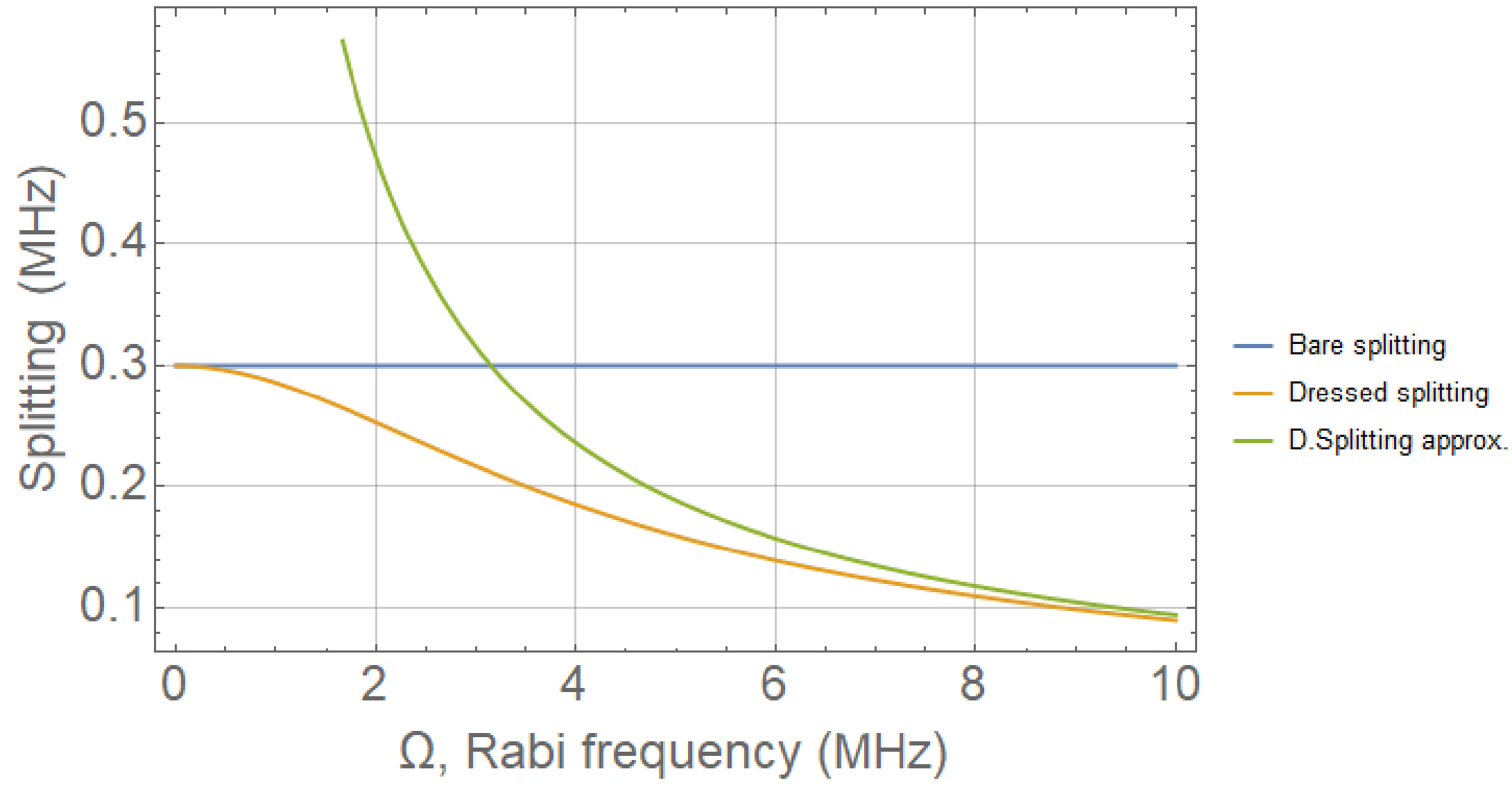

426 Fig. S8. Comparison of the energy splitting between the bare qubit and single-dressed qubit.
427 The green curve (dressed splitting approximation) is based on eq. S1. $A = 0.3$ MHz, and $\delta = $
428 3 MHz.
429
430
431
432
433

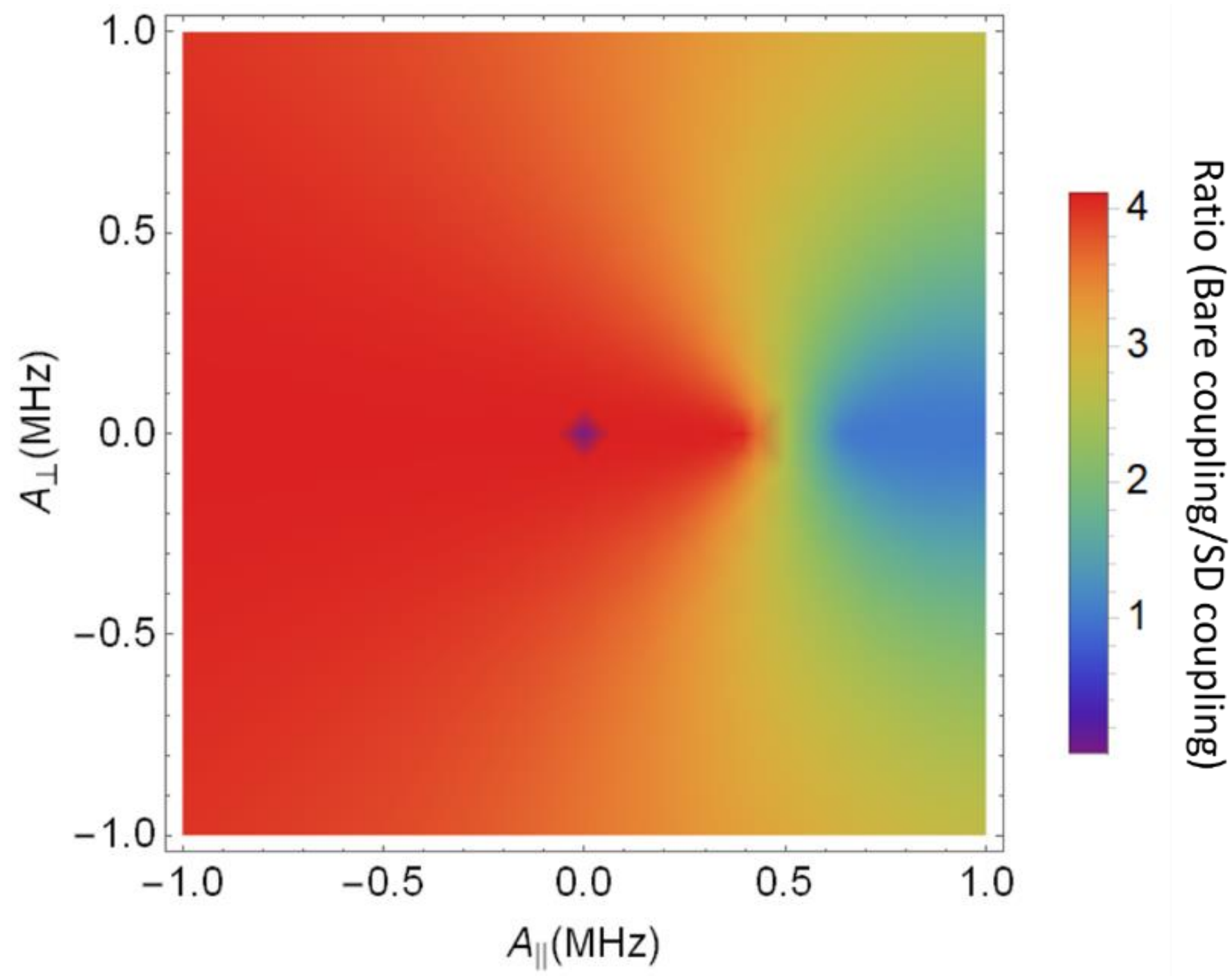

Fig. S9. Comparison of the hyperfine splitting between the bare spin qubit and the single-dressed qubit. We assume 405 G external field, 16 MHz Rabi frequency, and 4 MHz detuning. The minimum occurs around $A_{||} \sim 0.867$ MHz, where $A_{||}$ equals to $2\gamma_{\mathrm{n}}B$ i.e. twice of the $^{13}$C Larmor frequency at 405 G.

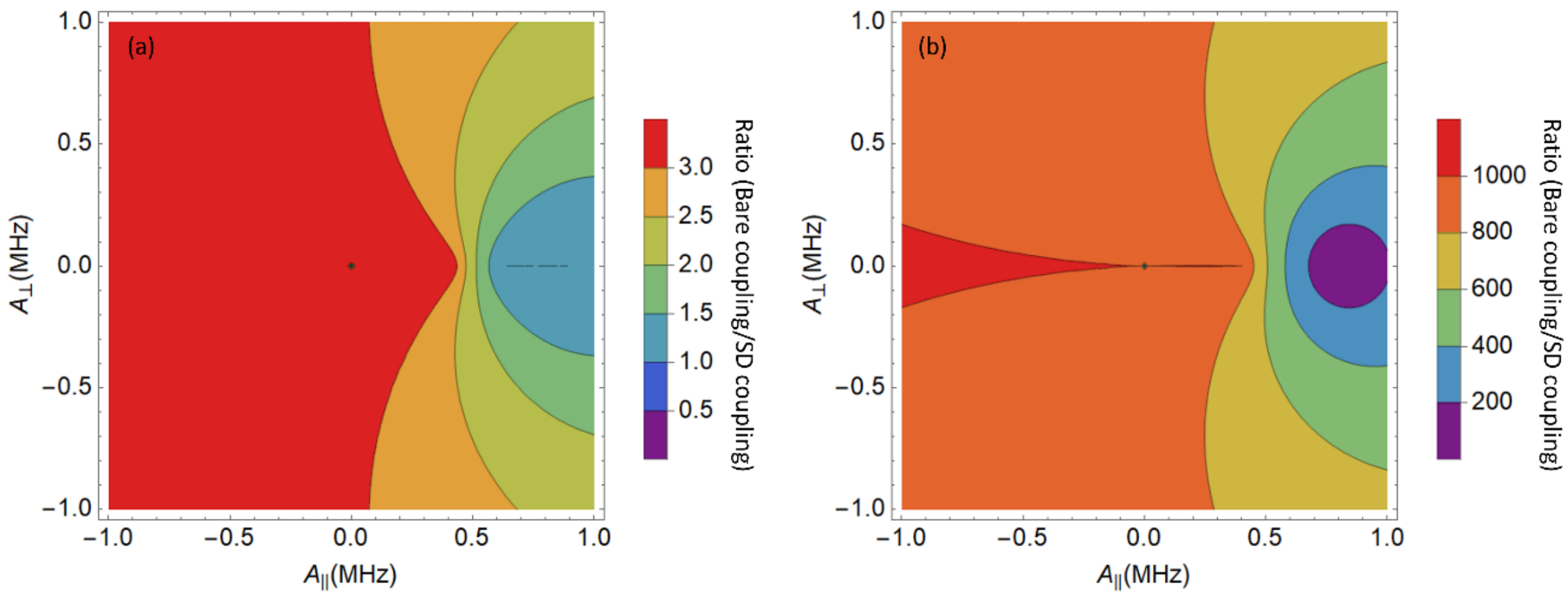

Fig. S10. Numerically calculated hyperfine splitting with different detuning regimes. For better
visualization, contour maps are used. (a) 3 MHz detuning and 10 MHz Rabi frequency (b) 0.01
MHz detuning and 10 MHz Rabi frequency.

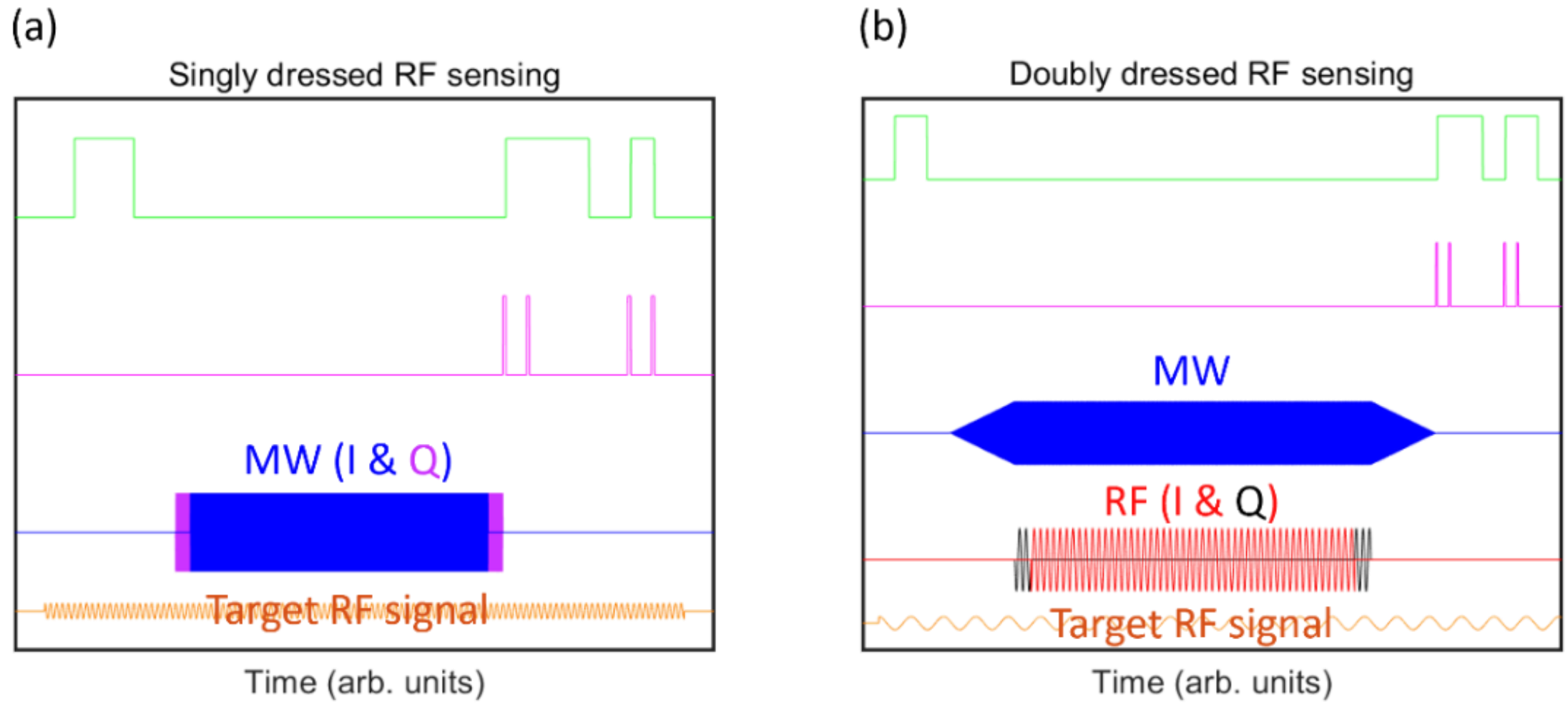

Fig. S11. Pulse sequences for dressed AC magnetometry. (a) single-dressed and (b) double-
dressed. The green solid lines are green laser pulses, and the purple lines are APD readout
pulses.

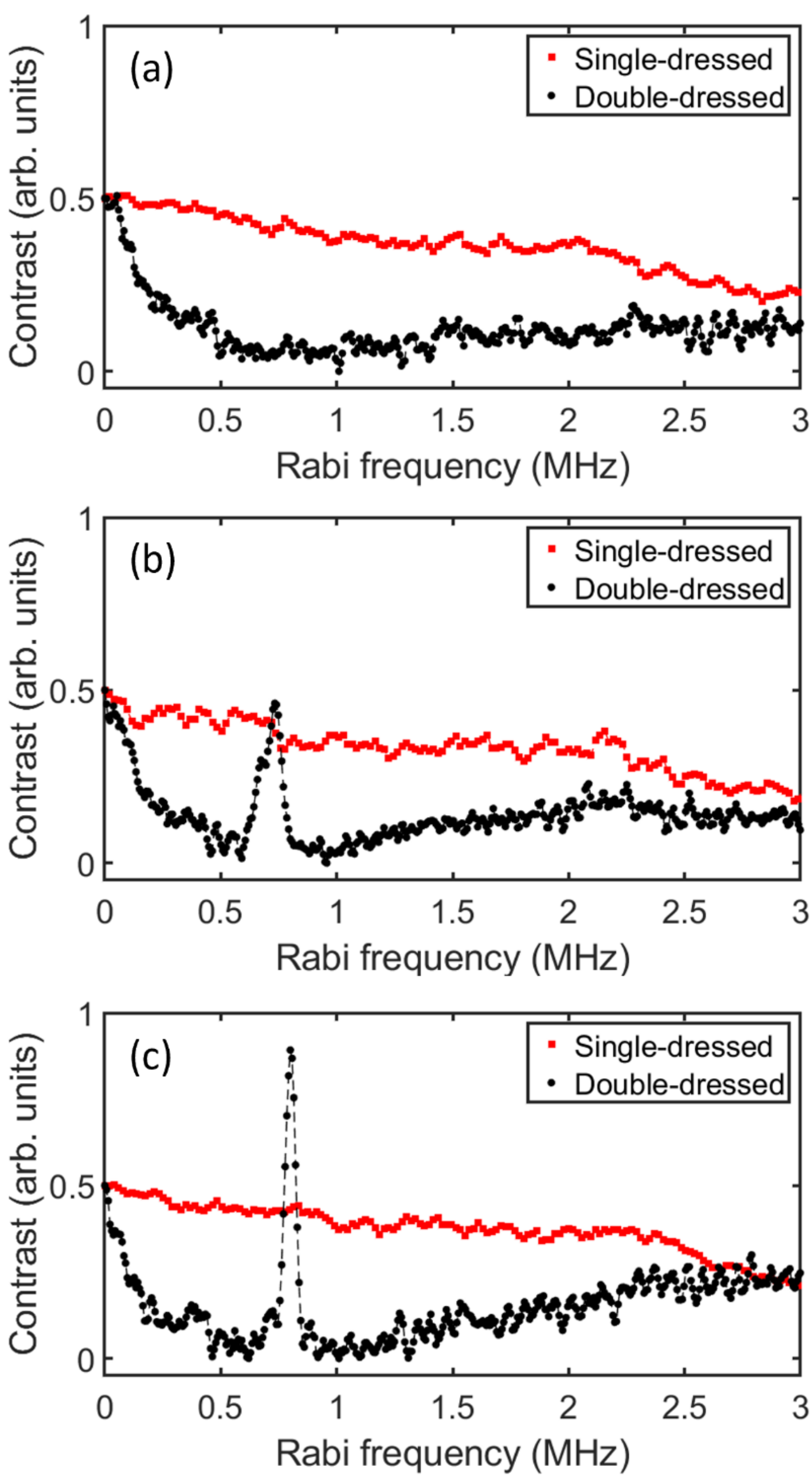

Fig. S12. Sub-MHz AC magnetometry with coherent AC signal. The applied magnetic field is
377 G. (a) With no applied signal, (b) with randomly phased stochastic AC signals at 0.8 MHz
and (c) with coheret 0.6 MHz AC signal. The noise floor is further reduced by up to 12 dB. In
(b), the phase averaged signal shows less amplitude and some broadening, skewing or shifting.
(c) A coherent AC magnetic field at 0.8 MHz with an amplitude of 3.17 μT is applied as a test
AC signal, which is detected with SNR ≈ 17 in the double-dressed measurement.s

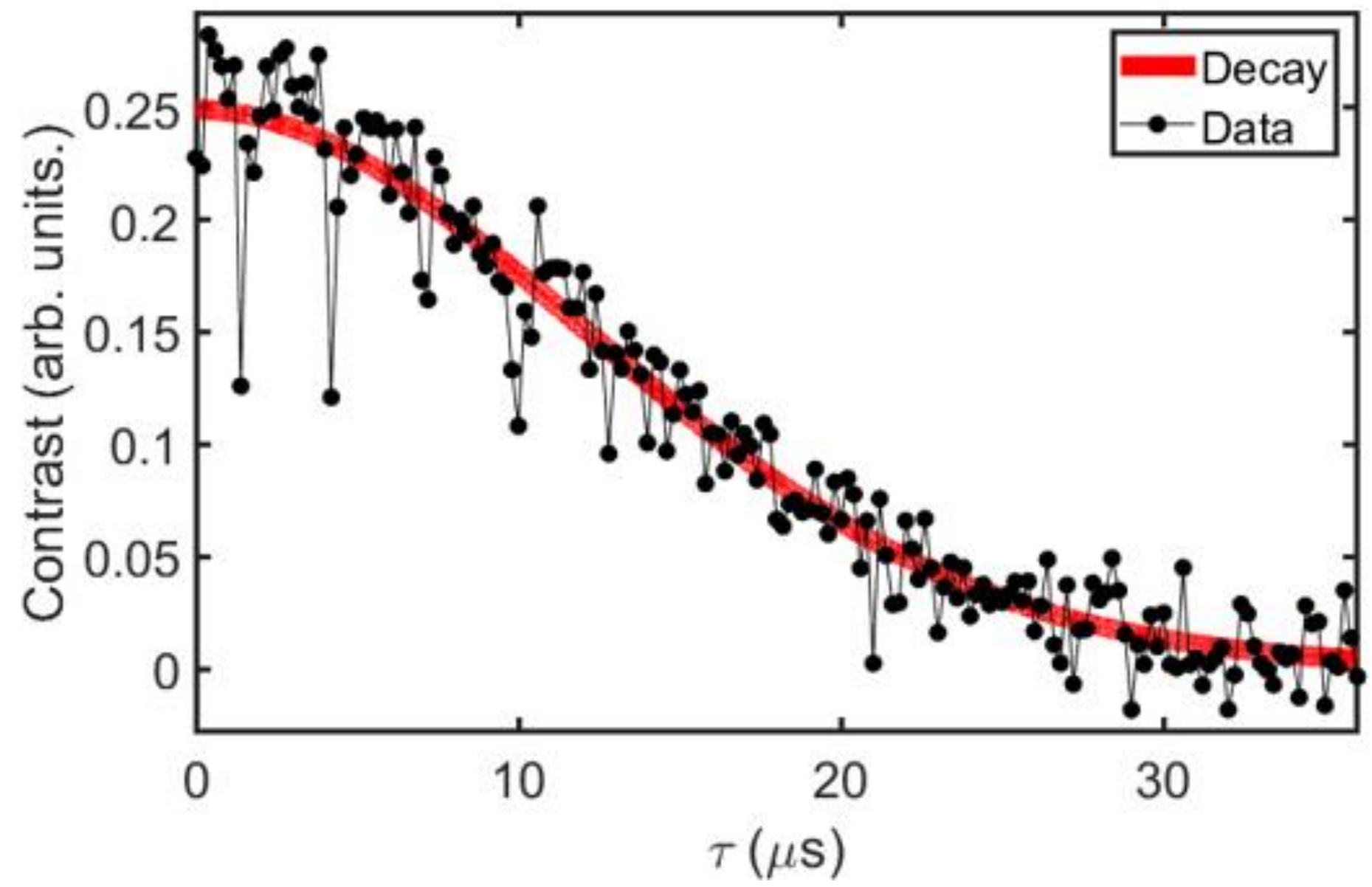

Fig. S13. Coherence decay acquired from XY16 experiment. The total evolution time is ($16\tau$ +
gate times). The $T_2$ coherence time without RF noise is $274.57 \pm 8.78$ μs.

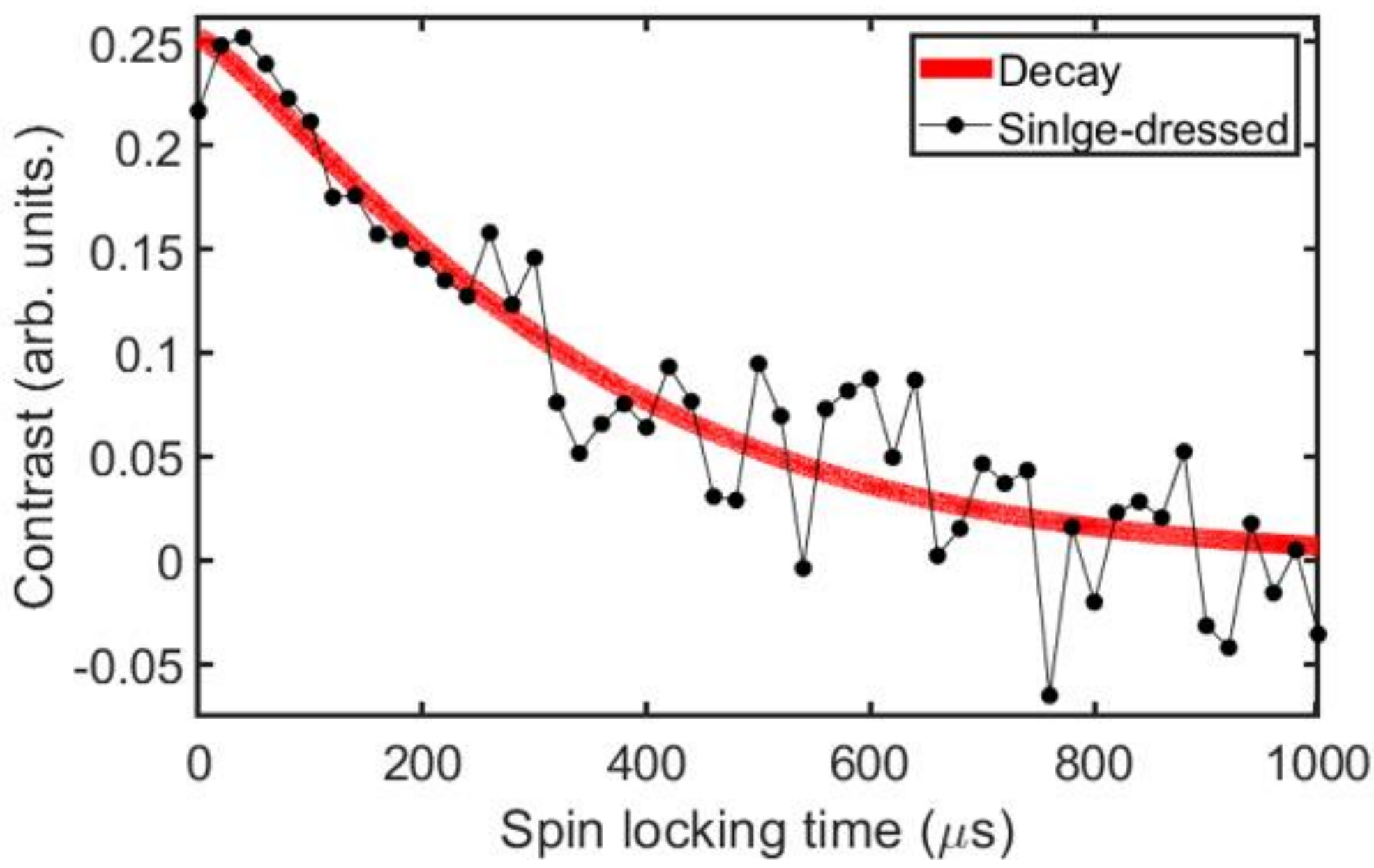

Fig. S14. Coherence decay acquired from single-dressed spin locking experiment. $T_{1,\rho}$
coherence time without RF noise is 346.84 $\pm$ 49.32 μs

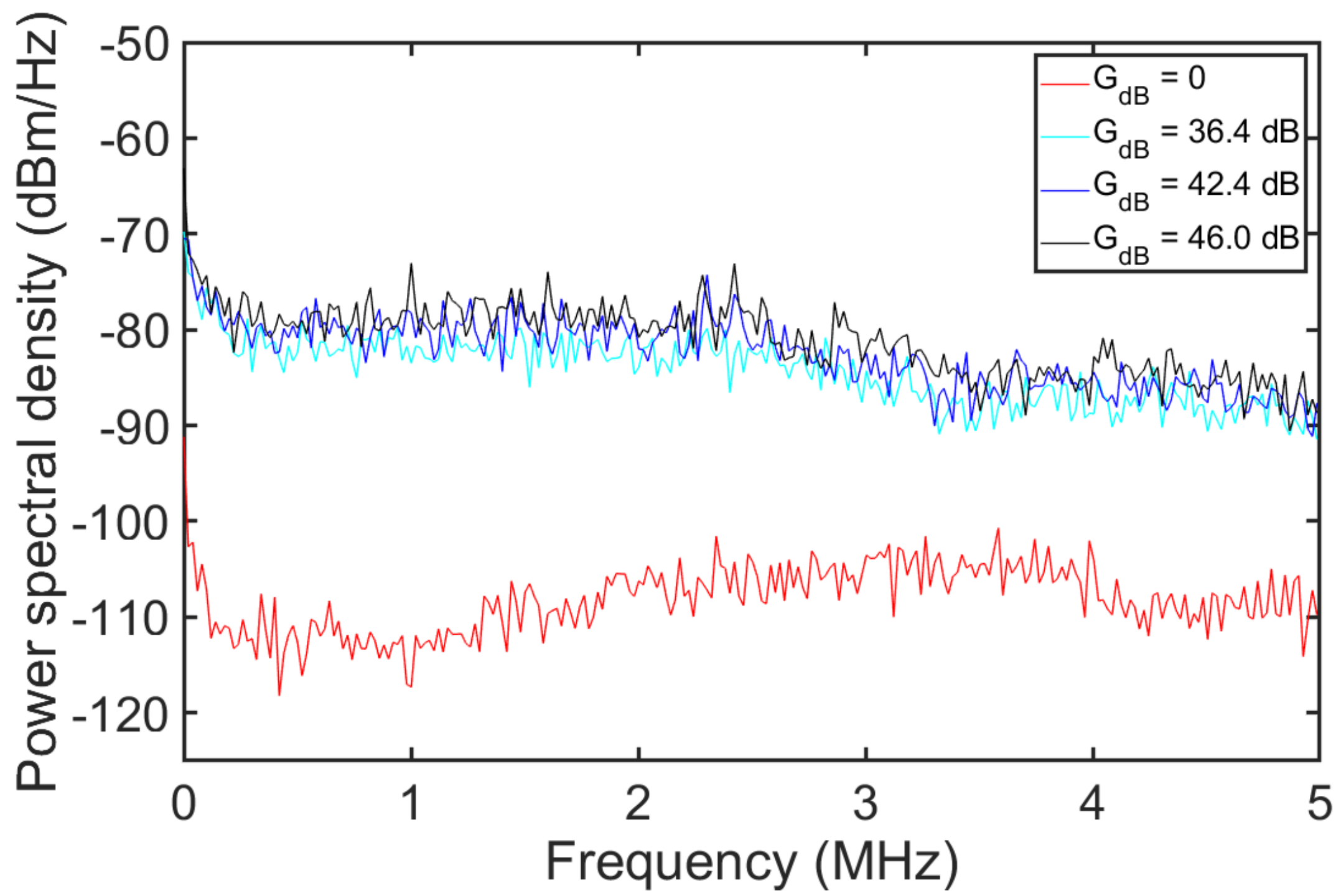

Fig. S15. Power spectral density (PSD) of the RF noise measured with an oscilloscope
(Keysight, DSAV134A). $G_{\mathrm{dB}}$ denotes the power gain of the RF amplifier at 1 MHz. The
integrated noise power ($P_{\mathrm{noise}}$) values, obtained by integrating the PSD from 0 Hz to 5 MHz,
are 9.22 nW, 56.06 nW, 71.67 nW, 123.30 nW for gain setting of 0, 36.4 dB, 42.4 dB and 46.0
dB, respectively.